\documentclass[aps,pre,twocolumn,superscriptaddress,showpacs,floatfix,nofootinbib,longbibliography]{revtex4-2}
\usepackage{dcolumn}
\usepackage{amsmath}
\usepackage{amssymb}
\usepackage{amsthm}
\usepackage{graphicx}
\usepackage{bm}
\usepackage[T1]{fontenc}
\usepackage{color}
\usepackage{url}
\usepackage[bf]{subfigure}
\usepackage{rotating}
\usepackage{scalefnt}
\usepackage{multirow}

\usepackage{scalefnt}
\usepackage{times}
\usepackage{mathtools}
\usepackage{bbm}
\usepackage{float}
\usepackage{enumerate}
\usepackage{algorithm}
\usepackage{algpseudocode}
\usepackage{hyperref}

\usepackage{float}
\usepackage{enumerate}

\begin{document}

\title{Assigning Entities to Teams as a Hypergraph Discovery Problem}

\author{Guilherme Ferraz de Arruda$^*$}
\affiliation{CENTAI Institute, Turin, Italy}

\author{Wan He$^*$}
\affiliation{Network Science Institute, Northeastern University, Boston, MA, USA}

\author{Nasimeh Heydaribeni$^*$}
\affiliation{Department of Electrical and Computer Engineering, University of California, San Diego}

\author{Tara Javidi}
\affiliation{Department of Electrical and Computer Engineering, University of California, San Diego}

\author{Yamir Moreno}
\affiliation{Institute for Biocomputation and Physics of Complex Systems (BIFI), University of Zaragoza, 50018 Zaragoza, Spain}
\affiliation{Department of Theoretical Physics, University of Zaragoza, 50018 Zaragoza, Spain}
\affiliation{CENTAI Institute, Turin, Italy}

\author{Tina Eliassi-Rad}
\affiliation{Network Science Institute, Northeastern University, Boston, MA, USA}
\affiliation{Khoury College of Computer Sciences, Northeastern University, Boston, MA, USA}
\affiliation{Santa Fe Institute, Santa Fe, NM, USA}

\begin{abstract}
We propose a team assignment algorithm based on a hypergraph approach focusing on resilience and diffusion optimization. Specifically, our method is based on optimizing the algebraic connectivity of the Laplacian matrix of an edge-dependent vertex-weighted hypergraph. We used constrained simulated annealing, where we constrained the effort agents can exert to perform a task and the minimum effort a task requires to be completed. We evaluated our methods in terms of the number of unsuccessful patches to drive our solution into the feasible region and the cost of patching. We showed that our formulation provides more robust solutions than the original data and the greedy approach. We hope that our methods motivate further research in applying hypergraphs to similar problems in different research areas and in exploring variations of our methods.
\end{abstract}

\maketitle
\def\thefootnote{*}\footnotetext{These authors contributed equally to this work}

\section{Introduction}

The Team Formation Problem (TFP) consists of finding assignments of individuals to one or more tasks. The optimal solution often depends on a subjective definition of the fitness function~\cite{juarez_comprehensive_2021}.
This problem was originally proposed in~\cite{Kuhn1955Hungarian} and has been studied from many different perspectives~\cite{Kuhn1955Hungarian, Abhilash2013, Okimoto2015, Okimoto2016M, Katsumi2018, Abhilash2013, Aijun2013, Evimaria2014, Qing2015, juarez_comprehensive_2021, KUSIAK1999, Fitzpatrick2005, Feng2010, Lin2004, Aijun2011, Lappas2009, juarez_comprehensive_2021, juarez_comprehensive_2021}. The most studied formulations include the formation of robust and recoverable teams~\cite{Abhilash2013, Okimoto2015, Okimoto2016M, Katsumi2018, juarez_comprehensive_2021}, budget and profit optimization~\cite{Abhilash2013, Aijun2013, Evimaria2014, Qing2015, juarez_comprehensive_2021}, and single or multi-skilled candidate optimization~\cite{KUSIAK1999, Fitzpatrick2005, Feng2010, Lin2004, Aijun2011, Lappas2009, juarez_comprehensive_2021}, among others. Regardless of the fitness function, the optimal solution often requires the computation of all possible assignments, which, due to the combinatorial nature of the problem, makes it
indeed NP-hard~\cite{juarez_comprehensive_2021}. Therefore, heuristics are often used to obtain locally optimal solutions that provide a tradeoff between computational cost and the quality of the solution. 

Solving the TFP is important beyond its theoretical and computational interest. Ref.~\cite{juarez_comprehensive_2021} also discusses modern applications of TFP, such as using the team formation problem to explore what-if scenarios. For example, a company could use the TFPs to draw possible future scenarios before hiring and also for candidate selection based on their ability to improve the space of possible teams for future tasks. Along similar lines, another application would be Labor Strategy Optimization~\cite{juarez_comprehensive_2021}, where TFPs could be useful to inform decisions about an organization's capability, location, and flexibility given a desired demand.

Moreover, the authors of the recent review~\cite{juarez_comprehensive_2021} propose the relationship between the TFP and the N-body problem in physics. The argument is based on the observation that the aggregation of a set of pairwise interactions does not capture the dynamics between groups of people. Interestingly, this is the same motivation for studying hypergraphs in complex systems~\cite{Iacopo2019, Jhun_2019, Arruda2020, Battiston2021, barrat2021social, Battiston2021b, Arruda2023, Boccaletti2023}. In this case, the motivation to use hypergraphs $-$or other forms of higher-order interactions$-$ is the analysis of dynamical phenomena, such as social contagion, in which we have one-to-many and/or many-to-many interaction types~\cite{Iacopo2019, Jhun_2019, Arruda2020, barrat2021social, Arruda2023}. In addition to this class of models, hypergraphs have been studied in a wide variety of problems, including cooperation in groups~\cite{Alvarez-Rodriguez2021} and percolation~\cite{Bradde_2009, bianconi2023theory}, showing that hypergraphs can be significantly different from graphs both in how individuals interact and in how fragile these structures can be. In particular, assuming that the failure of some nodes implies the failure of the hyperedge, the results in~\cite{bianconi2023theory} suggest that hypergraphs can be very fragile.

Here, we focus on a TFP in which teams are robust and recoverable and propose a hypergraph-based approach to perform the task assignments. In our context, robustness is defined as the ability of a task assignment to complete the tasks after removing an agent or a set of agents.
We also want to incorporate heterogeneity in (i) the importance of agents in the assigned tasks and (ii) the energy and budgets of tasks and agents. The rationale behind this choice is that an agent may have a fundamental role in one task but a less important one in another project.
As an example, consider scientific collaborations where a researcher might simultaneously lead one project and play a lesser role by contributing to other projects. This will also be reflected in the time the researcher spends on each project, which may have different requirements. Therefore, we propose to use edge-dependent vertex weight (EDVW) hypergraphs~\cite{Chitra2019}, where agents (nodes) can have different weights in different tasks (groups or hyperedges), capturing the heterogeneities of agents and tasks. Moreover, inspired by the study of resilience in graphs~\cite{Jamakovic2008, Cozzo2019}, we propose to use the algebraic connectivity of the Laplacian matrix~\cite{Chitra2019}, which is an ideal candidate for summarizing the robustness, including all the information of an EDVW hypergraph. The advantage of this approach is that, as a by-product of optimizing robustness via algebraic connectivity, we simultaneously reduce the diffusion timescale, hopefully facilitating communication between agents. In addition, we explicitly propose to include the energies and budgets as constraints in our optimization algorithm, which also captures this type of heterogeneity.

We systematically analyzed algebraic connectivity in small hypergraphs to validate and understand its behavior. As a case study, we also evaluated a publication dataset. This dataset can be interpreted as a collaboration hypergraph, where publications are tasks and authors are agents. The results suggest that our approach captures more robust assignments, where the failure of some nodes does not imply the failure of the tasks. Using the publications in Physical Review E (PRE) from 1993 to 2021, we show that the optimized hypergraphs have lower patching costs in most cases. Also, the number of unsuccessful patches is practically zero, while it can be up to $60\%$ in the original non-optimized hypergraph. To summarize, our main contributions are:
\begin{itemize}
 \item We map the team formation problem as a hypergraph in Sec.~\ref{sec:Mapping};
 \item We systematically analyze some small hypergraph cases in sections~\ref{sec:Small} and~\ref{sec:percolation};
 \item We perform a finite-size analysis for some classes of hypergraphs in Sec.~\ref{sec:percolation};
 \item We propose a constrained simulated annealing approach to maximize the algebraic connectivity of the mapped hypergraph in Sec.~\ref{sec:CSA};
  \item As a baseline, we propose a greedy algorithm  to maximize the algebraic connectivity of the mapped hypergraph in Sec.~\ref{sec:exp};
  \item We measure the resilience of our solutions based on attacks in Sec.~\ref{sec:results};
 \item We evaluate performance on two real datasets about scientific collaborations, APS~\cite{APS_website} and MAG~\cite{sinha2015overview}. We show that, in most cases, optimized hypergraphs have lower patching. Also, the number of unsuccessful patches is virtually zero, while it can be up to $60\%$ in the original hypergraph. These results are shown in Sec.~\ref{sec:results};
 \item We compare our hypergraph formulation with a bipartite formulation in Sec.~\ref{sec:Bipartite}.
\end{itemize}

\section{Background Information and Related Work}
\label{sec:background}

\subsection{Hypergraphs, Random Walks, and the Laplacian Matrix}
\label{sec:Laplacian}

The edge-dependent vertex weighted hypergraph is defined as $\mathcal{H} = \{\mathcal{V}, \mathcal{E}, \omega, \gamma \}$, where $\mathcal{V} = \{v_1, v_2, \dots, v_N\}$ is the set of nodes, $\mathcal{E} = \{e_1, \dots, e_K\}$ is the set of hyperedges, which are subsets of nodes of arbitrary size, $\omega(e_k)$ is a function weighting the hyperedges, and $\gamma(v_i, e_k)$ is a function weighting the importance of node $v_i$ in hyperedge $e_k$. Note that a node can have different weights depending on the hyperedge. This type of hypergraph is particularly interesting for modeling rich data with context-dependent weights. Literature examples include collaboration networks~\cite{Chitra2019}, machine learning applications such as hypergraph neural networks~\cite{Feng2019, Yadati2019, Ji2020, HNHN2020, ijcai21-UniGNN, Gao2023, Li2023}, and chemical reactions~\cite{Jost2019, Mulas2021}, among many others.

We define the weighted degree of each agent $d(v_i) = \sum_{e \in \mathcal{E}(v_i)} \omega(e)$, and the weighted degree of each task $\delta(e) = \sum_{v \in e} \gamma(v, e)$. A random walk that captures all the relationships and weights in an EDVW hypergraph can be defined as a sequence of nodes where: (i) the walker in node $v_i$ chooses a hyperedge $e$ according to its weighted degree, i.e., $\frac{\omega(e)}{d(v_i)}$, next (ii) the walker chooses a node $v_j$ within hyperedge $e$ with probability proportional to its hyperedge degree, i.e., $\frac{\gamma(v_j, e)}{\delta(e)}$.

Next, we define the hyperedge weight matrix $W \in \mathbb{R}_+^{N \times K}$ whose components $W_{ik} = \omega(e_k)$ if node $v_i$ is in the hyperedge $e_k$ and $W_{ik} = 0$ otherwise; the degree matrix $D_V \in \mathbb{R}_+^{N \times N}$, a diagonal matrix whose components are the weighted degree of each agent, i.e., $[D_V]_{ii} = d(v_i)$; the hyperedge degree matrix $D_E \in \mathbb{R}_+^{K \times K}$ which is a diagonal matrix whose components are the weighted degree of each hyperedge, $[D_E]_{kk} = \delta(e_k)$; and the vertex-weights matrix as $R \in \mathbb{R}_+^{N \times K}$, whose components $R_{ik} = \gamma_{e_k} (v_i)$.
Thus, the probability transition matrix for our random walk is expressed as
\begin{equation}
 P = D_V^{-1} W D_E^{-1} R^T,
\end{equation}
where $P \in \mathbb{R}^{N \times N}$ is usually asymmetric.

We can now define the combinatorial Laplacian matrix as~\cite{Chitra2019}
\begin{equation} \label{eq:L}
 L^H = \Pi - \frac{\Pi P + P^T \Pi}{2},
\end{equation}
where $\Pi_{ii} = \pi_i$ is a diagonal matrix and $\pi_i$ is the stationary distribution (the left eigenvector of $P$, i.e., $\pi P = \pi$).
This Laplacian matrix was originally defined in~\cite{Chitra2019} and is based on the Laplacian definition for directed graphs in~\cite{Chung2005}.
In~\cite{Chitra2019}, the authors argued that although it is a $N \times N$ symmetric object, it captures the essence of higher-order interactions.

Complementarily, since a diffusion process on this hypergraph depends on the algebraic connectivity of the Laplacian, we also expect that these weighting functions could capture the relationships between agents and tasks, reflecting the resilience of our system.
Formally, a diffusion process is defined as
\begin{equation}
 \dfrac{d \bm{x}(t)}{dt} = -L^H \bm{x}(t),
\end{equation}
which solves as
\begin{align}
 \bm{x}(t) &= \exp \left( -L^H t\right) \bm x(0) \nonumber \\
  &= \sum_{i=1}^N \exp(-\mu_i t) \bm{v}_i \bm{v}_i^T\bm x(0) \nonumber \\
  &= \bm{v}_0 \bm{v}_0^T \bm x(0) + \exp(-\mu_2 t) \bm{v}_i \bm{v}_i^T\bm x(0) \nonumber \\
  & \qquad \qquad + \sum_{i = 3}^N \exp(-\mu_i t) \bm{v}_i \bm{v}_i^T\bm x(0) \nonumber,
\end{align}
where $\mu_i$'s are the eigenvalues of $L^H$ and $\bm{v}_i$ are their associated eigenvectors, and $0 = \mu_1 < \mu_2 < \cdots \leq \mu_N$.
Note that the algebraic connectivity defines the timescale of our process.
So, the larger the algebraic connectivity, $\mu_2$, the faster the diffusion.

\subsection{Team Formation Problems}
\label{sec:background_tfp}

The Team Formation Problem (TFP) is informally defined as the matching of team members or agents to form one or more teams.
This class of problems can be formulated in many ways, targeting different team characteristics and goals~\cite{Kuhn1955Hungarian, Abhilash2013, Okimoto2015, Okimoto2016M, Katsumi2018, Abhilash2013,  Aijun2013, Evimaria2014, Qing2015, juarez_comprehensive_2021, KUSIAK1999, Fitzpatrick2005, Feng2010, Lin2004, Aijun2011, Lappas2009, juarez_comprehensive_2021, juarez_comprehensive_2021}. This problem has been tackled independently by operations research and data mining. Each field designs models and solutions according to its source of information. For example, research operations approaches are often driven by the requirements and needs of the organization~\cite{juarez_comprehensive_2021}.
On the other hand, the data mining approach focuses on social network data and friendship ties~\cite{juarez_comprehensive_2021}. Interestingly, regardless of the approach, these formulations often require the exploration of all possible team compositions, which relates to complex combinatorial problems that are often NP-hard problems or are suggested to be NP-hard~\cite{juarez_comprehensive_2021}.

Despite the approach taken, some characteristics are desirable for a good TFP solution. In~\cite{Rehman2021}, the authors suggest that TFP solutions should include: (i) reducing communication costs, (ii) being resilient (e.g., with respect to the removal of an agent), (iii) reducing personnel costs, (iv) balancing workloads, and (v) incorporating unique experts, skills, and leaders.
We note that our approach does not cover a skill set for the agents, and this feature can be incorporated as an additional set of constraints but is left as future work. Nevertheless, our methods cover all other desirable features.

Following the taxonomy proposed in~\cite{juarez_comprehensive_2021}, we focus here on the class of assignment-based models with many team formations and no team positions. Specifically, our problem is very similar to equations (13) to (16) in~\cite{juarez_comprehensive_2021}. The main difference with respect to that formulation is that in~\cite{juarez_comprehensive_2021}, the teams have a budget that cannot be exceeded, while we formulate our optimization problem in terms of the ``energy'' tasks need to be completed, imposing a lower bound. Furthermore, in~\cite{juarez_comprehensive_2021}, the authors suggest using TFP models to analyze possible scenarios during the resource planning phase. Concrete examples would be (i) identifying staffing shortages, (ii) recognizing training costs, and (iii) assisting in the optimization of labor strategies. In this particular application, our method can be used to identify resilient issues within a particular deployment and help provide more robust deployments.

We also note that in addition to the TFP, the proposed methodology can be applied to different problems. For example, in finance, a bipartite graph or hypergraph has been used to model a system of banks and assets~\cite{Huang2013, Caccioli2014}. In this scenario, the what-if scenario analysis proposed in~\cite{juarez_comprehensive_2021} could be helpful for a bank to determine which assets to sell or buy. Minor modifications may be needed to adapt current methods to specific scenarios.

Finally, regarding the literature on robust teams, various strategies have been proposed in the literature~\cite{juarez_comprehensive_2021}. Compared to our approach, one of the main differences is that our objective function is based on a hypergraph-based function that captures the resilience of the system. In contrast, the other approaches model it indirectly by adding additional terms to the optimization function or by including additional constraints.

\section{Problem formulation and analysis}
\label{sec:Formulation}

\subsection{Problem Definition}

We consider the task assignment problem with $N$ agents in the set $\mathcal{N} = \{1, \dots, N\}$, which should be assigned to one or more tasks in the set $\mathcal{K} = \{1, \dots, K\}$. A task requires $E_k$ units of ``energy'' to complete, which can be time, money, or other resources. Each agent has a total of $B_i$ units of energy, which is allocated to a set of tasks. Task $k$ is assigned to agent $i$ with weight $\mathcal{B}_{ik}$, meaning that agent $i$ will spend $\mathcal{B}_{ik}$ units of energy to complete task $k$. We assume that $\mathcal{B}_{ik}$ are integers. The matrix $\mathcal{B}=(\mathcal{B}_{ik})_{i \in \mathcal{N}, k \in \mathcal{K}} \in \mathfrak{B} = \mathbb{N}^{N\times K}$ represents the assignments. We also define the binary matrix $\bm X\in \mathcal{X} =\{0,1\}^{N \times K}$, whose elements $\bm x_{ik} = 1$ if agent $i$ is assigned to task $k$ with positive weight and $\bm x_{ik} = 0$ otherwise. The total energy units of the agents are denoted by $B=\sum_{i \in \mathcal{N}}B_i$. We must have $B\geq \sum_{k \in \mathcal{K}}E_k$ to have a feasible solution. We assume that agents pay a cost $f(\mathcal{B})$ for the task assignment $\mathcal{B}$. Our goal is to choose an assignment $\mathcal{B}$ that minimizes the cost $f(\mathcal{B})$. Thus, the optimization problem can be formalized as
\begin{subequations} \label{eq:constraints}
\begin{align}
 \min_{\mathcal{B}\in  \mathfrak{B}} &   f(\mathcal{B})  \label{eq:constraints_a} \\
 s.t. &\quad  \sum_{i \in \mathcal{N}}  \mathcal{B}_{ik} \geq E_k & \forall k \in \mathcal{K} \label{eq:constraints_b}\\
 & \quad \sum_{k \in \mathcal{K}} \mathcal{B}_{ik} \leq B_i & \forall i \in \mathcal{N} \label{eq:constraints_c}
\end{align}
\end{subequations}

We want to optimize the resilience of the final team assignment given the set of constraints in eqs.~\eqref{eq:constraints_b} and~\eqref{eq:constraints_c}. In other words, we want to choose a cost function $f(\mathcal{B})$ in Eq.~\eqref{eq:constraints_a} to capture the resilience of the final configuration. We will see later that the negative value of the algebraic connectivity of the hypergraph associated with the assignment $\mathcal{B}$ can be an appropriate cost function to provide the resilience we are looking for in the final assignment.

\subsection{Mapping The Problem as a Hypergraph}
\label{sec:Mapping}

\begin{figure*}[t!]
    \includegraphics[width=0.75\linewidth]{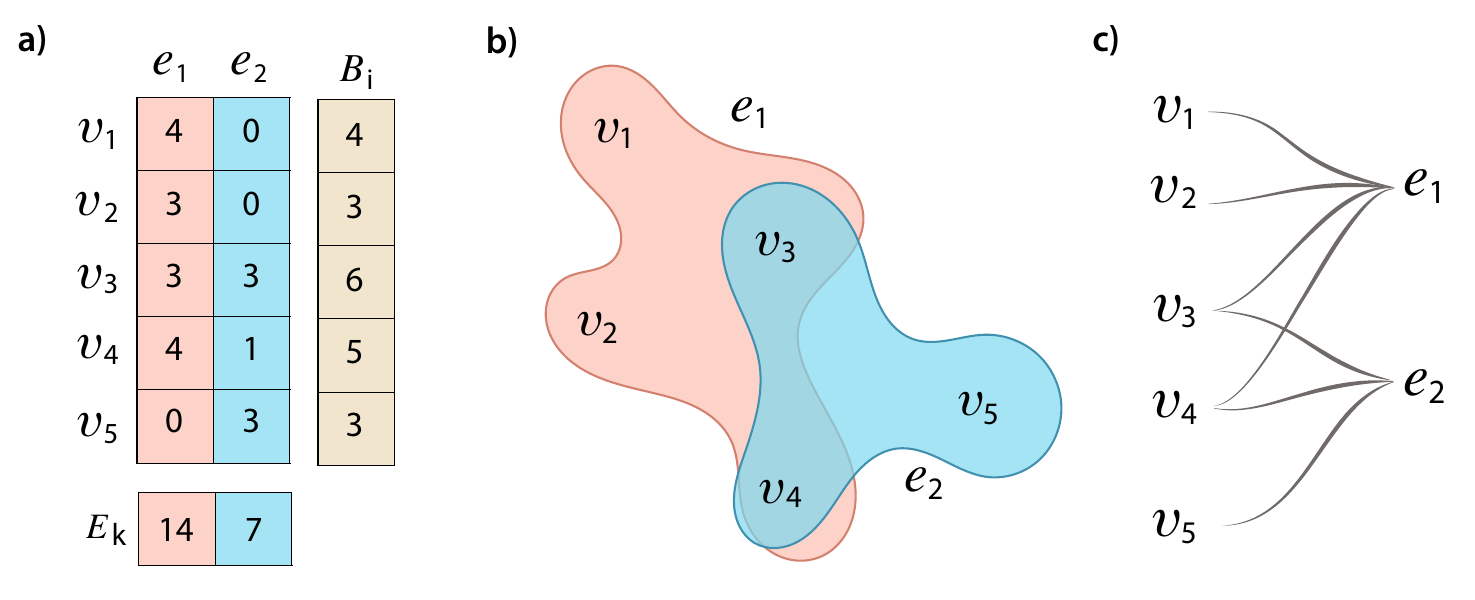}
        \caption{Graphical representation of the task-assignment problem. In (a), the task assignment is represented by $\mathcal{B}$ as well as an exemplary case of budgets, $B_i$'s, and energies, $E_k$'s, in (b) and (c), the hypergraph and bipartite representations of the same task assignment.}
    \label{fig:Schematic}
\end{figure*}

The problem formulated in Sec.~\ref{sec:Formulation} can be mapped into an edge-dependent vertex-weighted hypergraph, where we weigh both the hyperedges and the nodes within each hyperedge~\cite{Chitra2019}. Our optimization problem can be mapped to the EDVW hypergraph $\mathcal{H} = \{\mathcal{N}, \mathcal{E}, \omega, \gamma \}$, where $\mathcal{N}$ is the set of agents, $\mathcal{E}$ is the set of tasks (note that $|\mathcal{E}| = |\mathcal{K}|$ and the only difference between the sets $\mathcal{E}$ and $\mathcal{K}$ is the nature of the element in them). The weighting functions can be defined arbitrarily. Here, we propose to weigh the importance of a task as the energy required to complete it, i.e., $\omega(e_k) = E_k$. Also, the importance of an agent within a task is assumed to be the energy the agent spends in that task, $\gamma (v_i, e_k) = \mathcal{B}_{ik}$. Figure~\ref{fig:Schematic} (a) and (b) show this mapping graphically. Finally, we propose to maximize the algebraic connectivity of the Laplacian matrix in Eq.~\eqref{eq:L}, i.e., $\mu_2$, the second smallest eigenvalue of $L^H$.

The main advantage of formulating our problem as a hypergraph is that we can use the concepts of robustness and diffusion. By maximizing algebraic connectivity, we expect to make the hypergraph more resistant to attack and allow for faster diffusion processes. In practice, from the first, we hope that the failure of a node or task will have minimal impact. From the second, we expect the flow of information between agents to be as fast as possible. Note that from a TFP perspective, these are some of the desirable features for a solution. As mentioned in Sec.~\ref{sec:background_tfp}, in~\cite{Rehman2021}, the authors suggest that desirable features for TFPs include (i) reducing communication costs, (ii) being resilient (e.g., concerning the removal of an agent), (iii) reducing personnel costs, (iv) balancing workloads, and (iv) incorporating unique experts, skills, and leaders. We note that our proposed mapping focuses on a robust team assignment. However, algebraic connectivity maximization also reduces the timescale of the diffusion process, suggesting that communication costs are also reduced. Moreover, with respect to personnel costs and workload balancing, these features are incorporated into our method through the constraints. Thus, one can simply restrict the space of solutions to those that satisfy a given set of personnel costs and workload balance. We also note that we did not include different expertise and leaders in our formulation. In other words, all agents in the system can perform the tasks equally well. It should be noted that this assumption may be reasonable in some scenarios. Examples include the assignment of tasks to artificial agents, especially teams of robots~\cite{Liemhetcharat2012, Liemhetcharat2014, Okimoto2016M, juarez_comprehensive_2021}.

Here, we focus on connected hypergraphs. The reason for this choice is twofold. First, Laplacian matrices are semi-positive definite, so the multiplicity of zero eigenvalues is equal to the number of connected components in the hypergraph. Thus, if we optimize the algebraic connectivity in a hypergraph with multiple connected components, we can optimize only one component and neglect the others.
Second, from an application point of view, we want to increase communication between agents. In this case, we need to ensure that there is a path between any two agents.

We summarized our notation in Table~\ref{tab:notation}.

\begin{table}[]
\centering
\caption{Notation Summary}
\label{tab:notation}
\begin{tabular}{|c|c|}
\hline
Notation & Definition \\
\hline
$N$ & Number of agents \\
$K$ & Number of tasks \\
$\mathcal{N}$ & Set of agents \\
$\mathcal{K}$ & Set of tasks \\
$E_k$ & Energy requirement for task $k$ \\
$B_i$ & Energy budget of agent $i$ \\
$\mathcal{B}_{ik}$ & Energy exerted by agent $i$ towards completion of task $k$ \\
$\bm{X}$ & Assignment matrix \\
\hline
$\mathcal{H}$ & Edge-dependent, vertex-weighted (EDVW) hypergraph \\
$\mathcal{V}$ & Set of vertices (nodes) \\
$\mathcal{E}$ & Set of hyperedges \\
$\gamma(v_i, e_k)$ & Weight of vertex $v_i$ in hyperedge $e_k$ \\
$\omega(e_k)$ & Weight of hyperedge $e_k$ \\
$d(v_i)$ & Weighted degree of vertex $v_i$ \\
$\delta(e_k)$ & Weighted degree of hyperedge $e_k$ \\
$D_V$ & Vertex degree matrix \\
$D_E$ & Hyperedge degree matrix \\
$R$ & Vertex weight matrix \\
$W$ & Hyperedge weight matrix \\
$P$ & Probability transition matrix of the EDVW hypergraph $\mathcal{H}$ \\
$L^H$ & Laplacian matrix of the EDVW hypergraph $\mathcal{H}$ \\
$P^B$ & Probability transition matrix of the bipartite representation \\
$L^B$ & Laplacian matrix of the bipartite representation \\
$P^*$ & Probability transition matrix of the two-step bipartite  \\
$L^*$ & Laplacian matrix of the two-step bipartite  \\
$\mu_\ell(L)$ & $\ell$-th smallest eigenvalue of $L$ \\
$a$ & Scaling parameter, i.e., $\mu_2 \sim N^{-a}$ (context-dependent)\\
\hline
$\lambda_k$ & Penalty function for task $k$ \\
$\alpha$ & penalty function for the constraints \\
$\eta_i$ & Penalty function for agent $i$ \\
$\bar{T}$ & Average number of tasks assigned per agent \\
$\bar{A}$ & Average number of agents assigned per task \\
$\hat{A}$ & Average number of teammates per agent \\
\hline
\end{tabular}
\end{table}

\subsection{Small Hypergraph Examples}
\label{sec:Small}

\begin{figure*}[t!]
    \includegraphics[width=\linewidth]{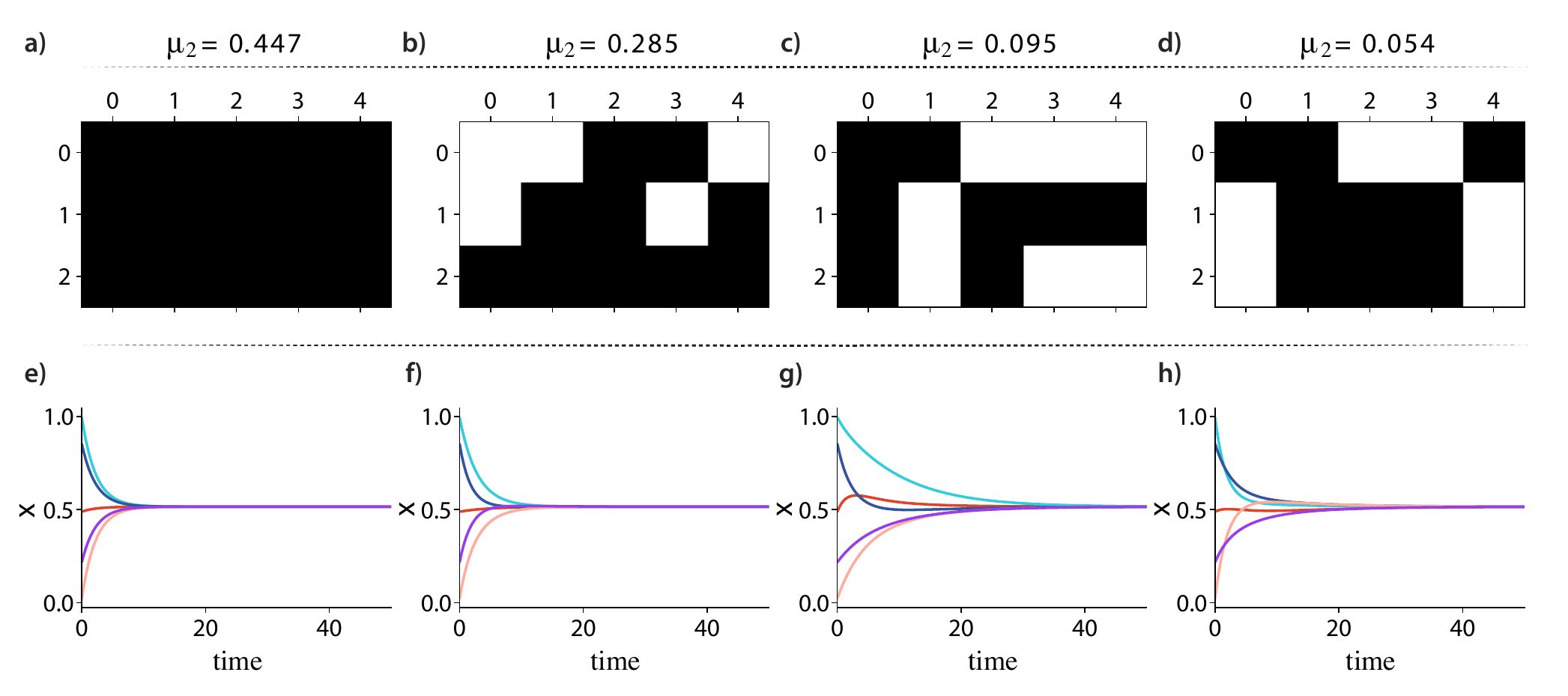}
        \caption{Examples of small hypergraphs. From left to right, in decreasing order of algebraic connectivity. In the top row, the graphical representation of the transposed incidence matrix is used for visualization, where the rows represent the hyperedges (tasks) and the columns represent the nodes (agents). The bottom row shows an example of the diffusion process defined by $L^H$. All processes start with the same initial condition, $\mathcal{B}_{ik} = 1$ for all assignments (see incidence matrices), $B_i =\sum_k \mathcal{B}_{ik}$, and $E_k =\sum_i \mathcal{B}_{ik}$.}
    \label{fig:SmallExamples}
\end{figure*}

To gain more insight into the behavior of optimizing algebraic connectivity, we focus on small hypergraphs where we can study the whole set of possible hypergraphs. We focus on hypergraphs with $N = 5$ and $K = 3$. We generate all possible hypergraphs with a single connected component, and the minimum cardinality is equal to or greater than two. We set $\mathcal{B}_{ik} = 1$ for all assignments in the generated hypergraph and $\mathcal{B}_{ik} = 0$ otherwise. The budgets and energies are defined as $B_i =\sum_k \mathcal{B}_{ik}$ and $E_k =\sum_i \mathcal{B}_{ik}$ to make the process as unconstrained as possible. We compute the algebraic connectivity for all these cases. In Fig.~\ref{fig:SmallExamples}, we show the highest and the lowest algebraic connectivity and two intermediate cases. In Fig.~\ref{fig:SmallExamples}(a) to (d), we show the graphically transposed incidence matrix, where the rows represent the hyperedges (tasks) and the columns represent the nodes (agents). In Fig.~\ref{fig:SmallExamples} (e) to (h), we show the behavior of a diffusion process on the studied hypergraphs. All hypergraphs have the same initial condition.

Fig.~\ref{fig:SmallExamples} shows that the higher the algebraic connectivity, the faster the diffusion. The analysis of this figure also suggests that unconstrained optimization of the algebraic connectivity favors denser hypergraphs. Nevertheless, we find that density does not fully explain intermediate cases. For example, Fig.~\ref{fig:SmallExamples} (c) is less dense than Fig.~\ref{fig:SmallExamples} (d) but has higher algebraic connectivity.
We note that the constrained problem will limit the space of possible solutions, providing an opposing force to the expansion favored by maximizing algebraic connectivity. In other words, we expect to be closer to the intermediate cases than to the bounded cases (Fig.~\ref{fig:SmallExamples} (a) and (d)).

\subsection{Exploring Hypergraph Structure and Algebraic Connectivity via Assignment Swapping}
\label{sec:percolation}

\begin{figure*}[t!]
\includegraphics[width=1\linewidth]{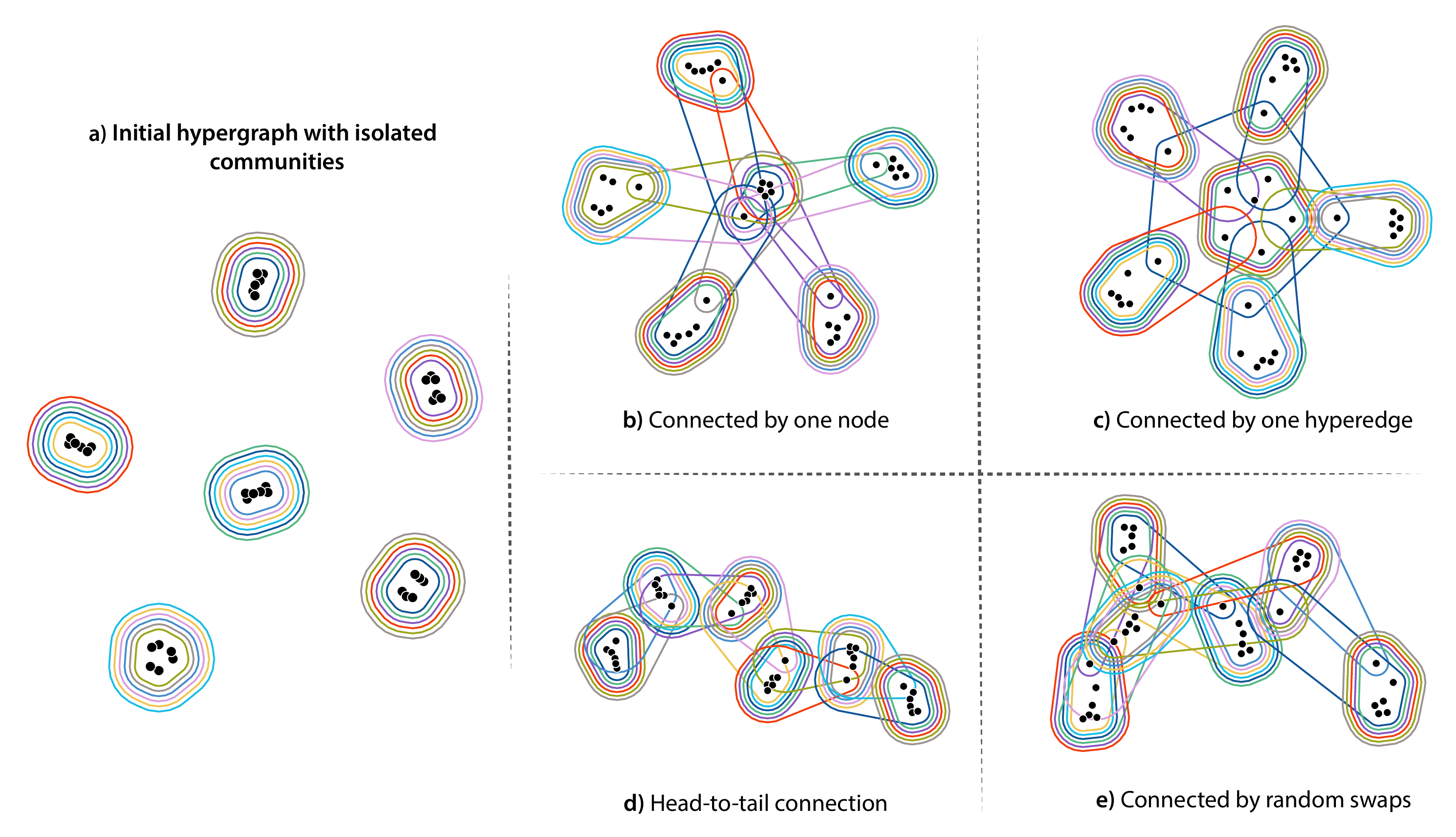}
\caption{Hypergraph assignment swapping with different structures, focusing on preserving constraints on agent budgets and task requirements. In this example, we start with a set of isolated communities, each formed by $6$ nodes sharing $6$ hyperedges, i.e., $6$ agents sharing $6$ tasks. We introduce different connections to these isolated communities with different structures to explore properties favorable to the algebraic connectivity function. For simplicity, we let $N_c$ be the number of communities in $\mathcal{H}$, where $n_i$ is the number of nodes in community $i$, and $m_i$ is the number of hyperedges in the community $i$.}
\label{fig:percolation}
\end{figure*}

\begin{figure}[t!]
\includegraphics[width=1\linewidth]{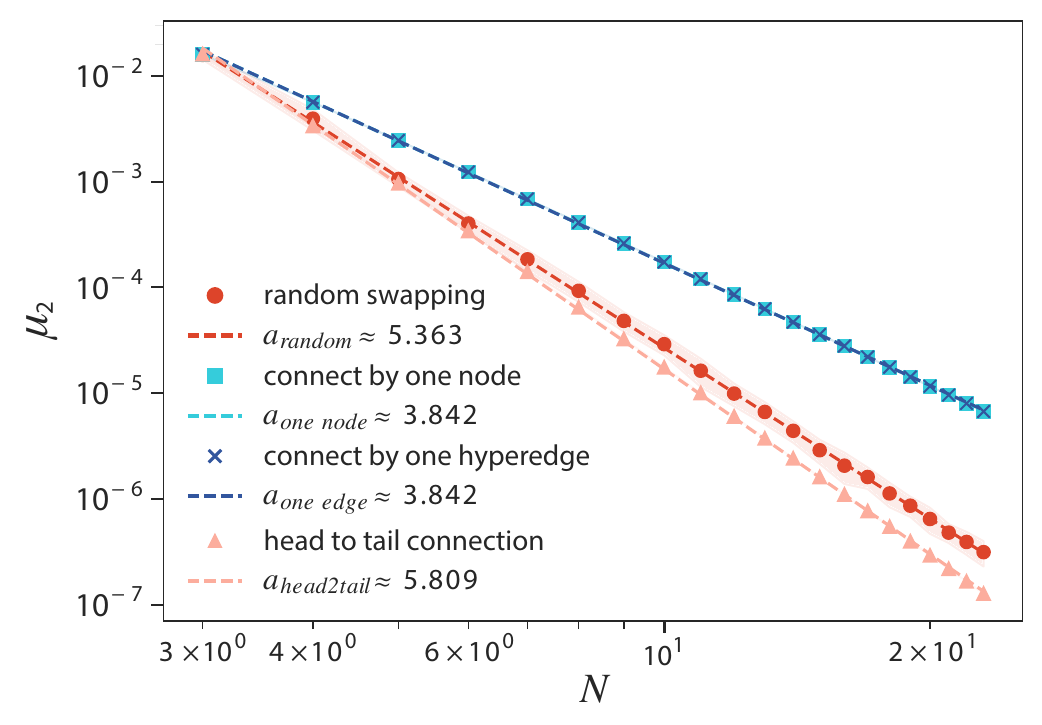}
\caption{Hypergraph assignment swapping on isolated communities. Here, we tested four different systems: (i) random, (ii) connected by one node, (iii) connected by one hyperedge, and (iv) head to tail (see Fig.~\ref{fig:Schematic}). We notice that the algebraic connectivity scales with the system size as $\mu_2\sim N_c^{-a}$, where, in this example, $N=n_c \times N_c, K= m_c \times N_c, N_c=n_{c_i}=m_{c_i}, \forall c_i \in \text{the set of communities}$. This figure is the result of $30$ independent repetitions.}
\label{fig:percolation_vs_size}
\end{figure}

Here, we want to understand the role of some representative small structures. We have designed a hypergraph percolation process and a set of experiments to investigate how different structures change the algebraic connectivity under an assignment-swapping setting. We start with a set of $N_C$ isolated communities, within which $m_{c_i}$ hyperedges are shared by the corresponding $n_{c_i}$ community members, and there is no communication between them.

We designed algorithms that create connections between communities with different emphases through assignment swapping while preserving edge and node weights. Since the edge and node weights preserve the character of the process, and assuming that the initial collaboration setup satisfies the set of constraints, each simulated hypergraph obtained through assignment swapping will satisfy the same set of constraints.

An analogy to the research collaboration experiment is to consider each community as a research lab or collaboration project. Initially, information was contained and could only flow freely within each community. Collaboration occurs only within the communities and rarely between the labs. Information diffusion between communities could occur after the introduction of system-wide information flows while satisfying the same set of constraints on agents' budget and task energy requirements by preserving node and edge weights. Analogously, inter-group collaborations or visiting research opportunities connect different communities and contribute to the flow of information, knowledge, and skills. We note that in the hypergraph setting, additional information flow could be achieved without adding additional nodes or hyperedges. Besides. swapping assignments within the hypergraph could introduce new connections, whether node-node, edge-node, or edge-edge, without sacrificing existing connections.

Figure~\ref{fig:percolation} shows a graphical example of the different types of hypergraphs analyzed. Given the initial setting, additional information flow could be facilitated by inter-community assignment swaps, which introduce collaboration between members of different communities. The more inter-community assignment swaps, the better the information diffusion between communities. Therefore, to fairly compare the swapping processes with different connection patterns and to evaluate their impact on algebraic connectivity, the number of inter-community assignment swaps is constant for each rewiring type. For simplicity, we let $N_c$ be the number of communities in $\mathcal{H}$, where $n_i$ is the number of nodes in the community $i$, and $m_i$ is the number of hyperedges in the community $i$. In a hub-like hypergraph connected by a single node, a centroid node is chosen to exchange with a random node in each of the other communities, and the end result would be a graph where all communities are connected through the centroid node as it participates in an edge in each community. Similarly, we can define a hub-like hypergraph connected by a single hyperedge. In this case, a centroid edge is chosen, and each community sends a node to join the centroid edge. Alternatively, a head-to-tail hypergraph can be formed by connections between each pair of consecutive communities. Finally, we also considered the case of random swaps, where connections between communities are formed by randomly swapping assignments between communities. 

Figure~\ref{fig:percolation_vs_size} shows the behavior of the algebraic connectivity as a function of the number of nodes. The hub-like hypergraph, i.e., connection through a node or an edge, consistently outperforms the other connection schemes. Note that in this setting, the connection through an edge and the connection through a node are identical due to the graph isomorphism to the hypergraph transpose. The linear head-to-tail hypergraph has the smallest algebraic connectivity. The randomized connection outperforms the linear head-to-tail hypergraph on average. We note that we observe a positive correlation between the algebraic connectivity and the average shortest distance. We also observe a power-law relationship between the algebraic connectivity and the size of the hypergraphs, where
\begin{equation*}
    \mu_2\sim N_c^{-a}, a>0,
\end{equation*}
and
\begin{equation*}
    a_{head2tail}>a_{random}>a_{one\ edge}=a_{one\ node}.
\end{equation*}
The number of introduced connections between communities is the same for graphs of the same size but differently connected for a fair algebraic connectivity comparison.

We should also note that the results in Fig.~\ref{fig:percolation_vs_size} also show an unusually high value for the exponents. This suggests that finite size effects play a role. However, due to the computational cost, we are unable to increase the system size to analyze large system sizes ($> 10^3$). Therefore, we decided to focus on small systems for this experiment. We should also note that the order observed between the algebraic connectivity for different systems is the main result of this analysis.

For most of the collaboration networks, we expect both the average and the maximum author budget and task energy requirements to be much smaller than the number of tasks or authors, i.e., $\mathbb{E}[\Omega],\ max(\Omega),\ \mathbb{E}[\Gamma],\ max(\Gamma) \ll |\mathcal{K}|$ and $\mathbb{E}[\Omega],\ max(\Omega),\ \mathbb{E}[\Gamma],\ max(\Gamma) \ll |\mathcal{N}|$, which means that under the constraints of the author's budget and task requirements, the collaborative network that optimizes algebraic connectivity is not expected to be centralized by a node or an edge, and is most likely to have a decentralized structure.

\section{The optimization problem}
\label{sec:OPT}

Motivated by applications in projects where tasks are interconnected and information diffusion between agents is an important factor, we propose to maximize the algebraic connectivity of $L^H$ as a quality metric. The rationale behind this measure is that higher-order interactions are well captured by the Laplacian. More specifically, they are captured by the algebraic connectivity of $L^H$. At the same time, by maximizing the algebraic connectivity, we simultaneously optimize robustness and information flow. However, the algebraic connectivity alone could drive the optimization algorithm to solutions where the agents are overworked. This problem is avoided by constraining the solution space. Thus, we rewrite the problem in Eq.~\eqref{eq:constraints} as\begin{subequations} \label{eq:constraintsL}
\begin{align}
 \max_{\mathcal{B}\in \mathfrak{B}} & \ \ \mu_2 (L^H)  \label{eq:constraintsL_a} \\
 s.t. & \sum_{i \in \mathcal{N}} \mathcal{B}_{ik} \geq E_k & \forall k \in \mathcal{K} \label{eq:constraintsL_b}\\
 & \sum_{k \in \mathcal{K}} \mathcal{B}_{ik} \leq B_i & \forall i \in \mathcal{N} \label{eq:constraintsL_c}
\end{align}
\end{subequations}

\subsection{Constrained Simulated Annealing}
\label{sec:CSA}

One of the approaches we use to solve the optimization problem~\eqref{eq:constraintsL} is constrained simulated annealing based on the penalty method. More specifically, we add a penalty function to our objective to penalize infeasible solutions generated by the simulated annealing optimization. The penalty function corresponding to the constraints in~\eqref{eq:constraintsL_b} is $-\lambda_k(E_k- \sum_{i \in \mathcal{N}}\mathcal{B}_{ik})^+$ for all $k \in \mathcal{K}$, and the penalty function corresponding to the constraints in~\eqref{eq:constraintsL_c} is
$-\eta_i(\sum_{k \in \mathcal{K}}\mathcal{B}_{ik} - B_i)^+$ for all $i \in \mathcal{N}$, where $(\alpha)^+=\alpha$ if $\alpha\geq0$ and $(\alpha)^+=0$ if $\alpha<0$. We keep the weights $\lambda_k$ and $\gamma_i$ fixed throughout the optimization. This method is implemented in Alg.~\ref{alg:CSA}.

We observed that to maximize the algebraic connectivity of the hypergraph, it is always best to use up all of the agents' budgets. Therefore, we initialize $\bm X$ so that $\sum_{k \in \mathcal{K}}x_{ik}=B_i$. Then, to use simulated annealing to maximize the algebraic connectivity, one only needs to swap the energy units in $\bm X$ (subtract and add energy units to and from $x_{ik}$'s) to generate new perturbations.

We also found that due to the search space being huge compared to the feasible region, adding the penalty function is insufficient to guide the simulated annealing algorithm to the feasible region. Therefore, we used a guided perturbation approach to push the samples toward the feasible region. The guided perturbation method is done by swapping the energy units in the incidence matrix so that we have less number of constraint violations after each round. The swapping of energy units is done according to two subroutines, which are chosen randomly (with adjustable probabilities). In the first subroutine, we randomly choose a task with extra energy, where the tasks with more extra energy are more likely to be chosen, and one energy unit of its assigned agents is transferred to a task that needs more energy. The second task is also randomly selected, with the tasks with more energy shortage being more likely to be selected. In the second subroutine, we randomly swap the energy units of the agents between two random tasks.

\begin{figure}[t!]
\begin{algorithm}[H]
\caption{Constrained Simulated Annealing (CSA)}
\label{alg:CSA}
\begin{algorithmic}
 \State \textbf{Input}:  Energy Requirements $E$, Budget Constraints $B$, The Initial Assignment $\mathcal{B}^0$. Optimization Parameters: $T^0$, $a_c$, $T_{th}$, $t_{max}$.
 \State Set the temperature $T=T^0$.
 \State Set $t=0$.
 \While{$T>T_{th}$ or $t<t_{max}$}
 \State Evaluate $\mathcal{B}^t$ and get penalty $P^t$, and $\tilde{E}^t$ from Alg.~\ref{alg:eval}.
\State Perturb $\mathcal{B}^t$ according to Alg.~\ref{alg:perturb} and get $\mathcal{B}^{t+1}$.
\State $t=t+1$
\EndWhile
\end{algorithmic}
\end{algorithm}
\end{figure}

\begin{figure}[t!]
\begin{algorithm}[H]
    \caption{CSA: Assignment Evaluation}
    \label{alg:eval}
    \begin{algorithmic}
   \State Input: $\mathcal{B}$, , $E$, $B$, $\eta$, $\lambda$.
    \State Output: $P$, $\tilde{E}$.
    \State
Calculate the algebraic connectivity, $e$, for the assignment $\mathcal{B}$.
\State Calculate penalty function: \\ $P= e - \eta_i(\sum_{k \in \mathcal{K}}\mathcal{B}^t_{ik} - B_i)^+ -\lambda_k(E_k- \sum_{i \in \mathcal{N}} \mathcal{B}^t_{ik})^+ $
\State Calculate constraint violations: \\
Set $\tilde{E}_{k}$= extra required energy for task $k$ ($\tilde{E}_{k}$ is negative if task $k$ has more than enough energy assigned to it)

\Return $P$ and $\tilde{E}$.
\end{algorithmic}
\end{algorithm}
\end{figure}

\begin{figure}[t!]
\begin{algorithm}[H]
    \caption{CSA: Assignment Perturbation}
    \label{alg:perturb}
    \begin{algorithmic}
        \State \textbf{Input} $\mathcal{B}$, $\tilde{E}$,  $N_s$.
        \For {$s= 1: N_s$}
\State Find tasks that require more energy, $H_p=\{k: \tilde{E}_k>0\}$.
\State Find tasks that have more than enough energy assigned to them, $H_n=\{k: \tilde{E}_k<0\}$.
\If {$H_p!=\emptyset$ and $H_n!=\emptyset$}
    \State Choose a task $h_p$ from $H_p$ with probabilities proportional to $\tilde{E}$.
\State Choose a task $h_n$ from $H_n$ with probabilities proportional to $|\tilde{E}|$.
\State Assign one energy unit of an agent assigned to task $h_n$ to task $h_p$.
\State Update $\mathcal{B}$ and $\tilde{E}$.
\Else
\State Choose two tasks randomly.
\State Choose one agent from each task randomly. Swap one of their energy units assigned to the chosen tasks with each other.
\State Update $\mathcal{B}$ and $\tilde{E}$.
\EndIf
        \EndFor
    \end{algorithmic}
\end{algorithm}
\end{figure}
The computational complexity of Algorithm \ref{alg:CSA} is determined by the computation of the algebraic connectivity, which is the second smallest eigenvalue of the Laplacian matrix ($N\times N$). The complexity of such a computation is $O(N^3)$. In each round of the CSA algorithm, the algebraic connectivity is computed once. Since the total number of rounds is bounded from above by a constant number, the total computational complexity of CSA is $O(N^3)$.

\subsection{Other Important Factors}

Although we chose to optimize the algebraic connectivity of the hypergraphs to get robust solutions, there are some other important factors to consider when evaluating a particular solution. The factors we considered are the average number of tasks an agent is assigned to, $\bar{T}$, and the average 
number of teammates an agent has, $\hat{A}$. It is not desirable to have solutions with large $\bar{T}$ and $\hat{A}$ because coordination and collaboration become more difficult as these factors increase. Therefore, we evaluate our solutions from this perspective and try to generate solutions with controlled levels of $\bar{T}$ and $\hat{A}$. We try two approaches to optimize the algebraic connectivity while having a controlled level of these quantities. The first approach is to penalize the objective function as these quantities increase. The second approach is to assign energy units to tasks in packs. That is, we can only assign $nP$ energy units of each agent to a task, where $P$ is the size of the energy pack and $n$ is an integer.

\section{Experiments}
\label{sec:exp}

\subsection{Datasets}

\label{sec:Datasets}

We focused on two collaboration datasets, the Microsoft Academic Graph (MAG)~\cite{sinha2015overview}, and the  American Physical Society (APS)~\cite{APS_website} datasets. The Microsoft Academic Graph (MAG) contains scientific publication records, citations, and other information. More information is described on MAG's website~\footnote{\url{https://www.microsoft.com/en-us/research/project/microsoft-academic-graph/}}.
The Collaborative Archive \& Data Research Environment (CADRE) project at Indiana University~\citep{mabry2020cadre} provided MAG's raw data. From the MAG dataset, we filtered the papers with the word "hypergraph" in their title and extracted the giant connected component of the authorship hypergraph (authors as nodes and papers as hyperedges).

The American Physical Society (APS) dataset contains the basic metadata of all APS journal articles from 1993 to 2021. From the APS dataset, we considered one journal (Physical Review E (PRE)), divided the dataset into 2-year intervals, and extracted the giant connected component of the authorship hypergraphs. We then optimized each of the extracted hypergraphs using the CSA and the greedy approach (described in the following).

\subsection{Baseline: The Greedy Approach}

A greedy optimization approach was implemented as a baseline solution. Inspired by the results presented in~\ref{sec:percolation}, which show that more centralized systems are favored when optimizing algebraic connectivity, the greedy algorithm starts with an initial hypergraph assignment that connects all tasks with the minimum number of hubs, where the hubs correspond to the highest-budget agents. These hubs are then connected by adding shared tasks among these highest-budget agents.

The greedy optimization can be divided into 2 phases. In phase 1, we start from the centralized initial assignment and then assign agents to tasks by filling the tasks with the agents that could lead to the maximum increase in the objective function per unit input energy until each task is full. In phase 2, we further optimize the objective by using up all the energy left in the authors after the assignment given in phase 1. To do this, we start with the assignment computed by phase 1 that satisfies the task completion requirement. Then, analogous to phase 1, each agent is assigned to the tasks that result in the most increase in goal per unit input until the agent's budget is exhausted. If the total task energy requirements are equal to the total agent budgets, there would be no excess agent budget available if all task requirements were met. As a result, there would be no need for phase 2 operation.

Regarding the suitability of applying simulated annealing to the greedy approach, since a viable solution has already been obtained in phase 1, it could be specified in phase 2 of the optimization whether the greedy optimization should be performed by rejecting assignments that lead to a negative change in the objective function or accepting such assignments with a probability.

The computational complexity of the greedy approach is higher than that of the CSA algorithm we use. The reason is that in each round of assignment in the greedy approach, the algebraic connectivity order must be computed $N$ times (the number of agents with available budget). The total number of rounds is of the order of the total number of tasks. Assuming that the number of tasks does not grow with $N$, the computational complexity of the greedy approach is $O(N^4)$.

\begin{figure}[t!]
    \begin{algorithm}[H]
    \label{alg:greedy_phase1}
        \caption{Greedy Knapsack Phase 1: Task Fulfillment}
        \begin{algorithmic}[1]
            \State \textbf{Input} $E$, $B$, $h$.
            \State \textbf{Output} $\mathcal{B}$.
            \State $h \in \mathbb{R}:$ The energy packet size that each energy assignment must be multiples of.\\
            \State Start with an empty initial assignment $\mathcal{B}^0$
            \While {Tasks are not all fulfilled}
                \For {each unfulfilled task $i$}
                    \State Potential energy spent by agent $j$ on task $i$: $e_{ij}=min(B_{j},E_{i},h)$.
                    \State Compute change in the objective function per unit energy input by agent $j$ when assigning $e_{ij}$ to task $i$.
                    \State Assign task $i$ to the agent that results in the maximum increase in objective function per unit input of energy.
                    \State Update assignment $\mathcal{B}$.
                \EndFor
                \State Update the list of unfulfilled tasks.
            \EndWhile
        \end{algorithmic}
    \end{algorithm}
\end{figure}

\begin{figure}[t!]
    \begin{algorithm}[H]
    \label{alg:greedy_phase2}
        \caption{Greedy Knapsack Phase 2:\\ Improve algebraic connectivity by using up agent energy}
        \begin{algorithmic}
            \State \textbf{Input} $E$, $B$, $\mathcal{B}_{phase 1}$, $h$.
            \State Continue with the assignment $\mathcal{B}_{phase 1}$ optimized in phase 1
            \While {Agent budget is not all used up}
                \For {each available agent $j$}
                    \State Potential energy spent by agent $j$ on task $i$: $e_{ij}=min(B_{j},h)$.
                    \State Compute change in the objective function per unit energy input by agent $j$ when assigning $e_{ij}$ to task $i$.
                    \State Assign task $i$ to the agent that results in the maximum increase in objective function per unit input of energy.
                \EndFor
                \State Update the list of available agents.
            \EndWhile
        \end{algorithmic}
    \end{algorithm}
\end{figure}

\subsection{Evaluation Metrics}

To evaluate our optimized hypergraphs, we investigate their resilience against agent removal attacks. The evaluation is based on how easy (or possible) it is to patch the attacked hypergraph and make it a feasible solution again. The considered patching process is inspired by a real-world scenario where agents form teams to complete tasks. If an agent fails to complete its task assignments (agent removal), the remaining agents must step in to complete the tasks. We assume that this will happen as follows. The agents assigned to the uncompleted tasks will first try to compensate for the removed agent. If these agents do not have an extra budget, they will have to ask other agents to fill in. However, they can only communicate with their teammates. If those teammates do not have any available budgets, they will ask their other teammates, and so on. The more hops we have to take to find replacements for the removed agents to find a feasible solution, the less resilient the original solution is to node removal attacks. We define the cost of patching based on the number of hops one has to go in the hypergraph and whether or not a feasible solution was found. The amount of unsuccessful constraints after patching is also another quantity we are interested in. We use Patching Cost and Unsuccessful Constraints as our two evaluation metrics. We compute these resilience metrics for the original hypergraphs obtained from the real datasets and the optimized ones obtained by our optimization algorithms.

\subsection{Other Network Representations}

For the sake of completeness, we should also discuss the alternative representations of higher-order data. In this context, we can mention simplicial complexes and the bipartite representations~\cite{Jamakovic2008, Cozzo2019}. Simplicial complexes can be understood as a special hypergraph whose set of hyperedges is complete, i.e., all possible subsets of a hyperedge are also present. Although this is a very useful tool in topological data analysis, the mutual inclusion in our context implies unnecessary complications in terms of weights. Therefore, we will not consider this representation. With respect to the corresponding bipartite representation, we consider, for comparison, a procedure similar to that of our hypergraph approach. In this section, we present this formulation, present our experimental results using this formulation, and compare them with the results from the hypergraph representation.

\subsubsection{Bipartite Laplacian matrix}
\label{sec:Bipartite}

First, we need to represent each random walk step individually to have a similar random walk interpretation. In our previous formulation, Sec.~\ref{sec:Laplacian}, the walker follows a two-step process, first visiting a hyperedge and then a node. The adjacency and degree matrix for the bipartite representation of our task assignment problem is
\begin{align*}
   A^B &=
    \begin{bmatrix}
    \bm 0 & (W \circ \chi) \\
    (\mathcal{B} \circ \chi)^T & \bm 0
\end{bmatrix}
\hspace{3mm}\text{and }\hspace{3mm}
D^B =
    \begin{bmatrix}
    D_V &\bm 0 \\
    \bm 0 & D_E
\end{bmatrix}.
\end{align*}

Thus, the probability transition matrix will be
\begin{align*}
P^B &= \left( D^B \right)^{-1} A^B =
\begin{bmatrix}
    \bm 0 & D_V^{-1} \left( W \circ \chi \right) \\
    D_E^{-1} \left( \mathcal{B} \circ \chi \right)^T & \bm 0
\end{bmatrix}
\end{align*}
In this case, the random walk is no longer a two-step walk. Consequently, the walker alternates between nodes to hyperedges and vice versa.
Although this is a mathematically reasonable formulation, more is needed to justify it physically. Specifically, in our case, the walker would alternate between ``visiting'' agents and tasks.
Next, following the same procedure as before, the Laplacian matrix is defined as
\begin{align*}
L^B &= \Pi^B - \frac{\Pi^B P^B + (P^B)^T \Pi^B}{2},
\end{align*}
where $\Pi^B$ is the random walk's stationary distribution defined by $P^B$.

To highlight the differences between our original formulation and the bipartite case, we first evaluate the random walk given by even steps. To obtain such a random walk, we must use the probability transition matrix given by $\left( P^B \right)^2$, formally expressed by
\begin{align*}
 &P^* = \left( P^B \right)^2 = \\
 &=
\begin{bmatrix}
     D_V^{-1} \left( W \circ \chi \right) D_E^{-1} \left( \mathcal{B} \circ \chi \right)^T &\bm 0 \\
    \bm 0 &  D_E^{-1} \left( \mathcal{B} \circ \chi \right)^T D_V^{-1} \left( W \circ \chi \right)
\end{bmatrix}.
\end{align*}
Because of its diagonal block structure we observe that the spectra of $P^*$ are the union of the spectra of $D_V^{-1} \left( W \circ \chi \right) D_E^{- 1} \left( \mathcal{B} \circ \chi \right)^T$ and the spectra of $D_E^{-1} \left( \mathcal{B} \circ \chi \right)^T D_V^{-1} \left( W \circ \chi \right)$. Regarding the random walk, if the walker starts in one mode of the bipartite graph, it will always stay in that mode. More importantly, the random walk defined by the upper left block is the same random walk defined in Sec.~\ref{sec:Laplacian}. Thus, defining the Laplacian as before, we have
\begin{align*}
L^* &= \Pi^* - \frac{\Pi^* P^* + (P^*)^T \Pi^*}{2},
\end{align*}
which is still a block diagonal matrix. On the other hand, $L^B$ is an off-diagonal block matrix. Another way of noting the differences between the random walks defined on these matrices is to note that in $L^B$, we alternate between modes of the bipartite, while in $L^*$, we always stay in the same mode, depending only on the initial position of the walker.

From a computational point of view, computing $L^B$ should also imply a higher computational cost. Note that $L^H \in \mathbb{R}^{N \times N}$, while $L^B \in \mathbb{R}^{N+K \times N+K}$ and that the matrix $L^* \in \mathbb{R}^{N+K \times N+K}$ has the same dimensions as $L^B$. However, since it is a block diagonal matrix, we can decompose the problem into smaller problems, one for each block on the diagonal. The first is exactly $L^B$, while the second lower diagonal matrix represents the Laplacian defined over the hyperedges and has the dimension $K \times K$. Since we are interested in algebraic connectivity, the higher the dimension of the matrices, the higher the computational cost.

\subsection{Examining the Influence of Constraints on Hypergraph Algebraic Connectivity}

\begin{figure}[t!]
    \includegraphics[width=1\linewidth]{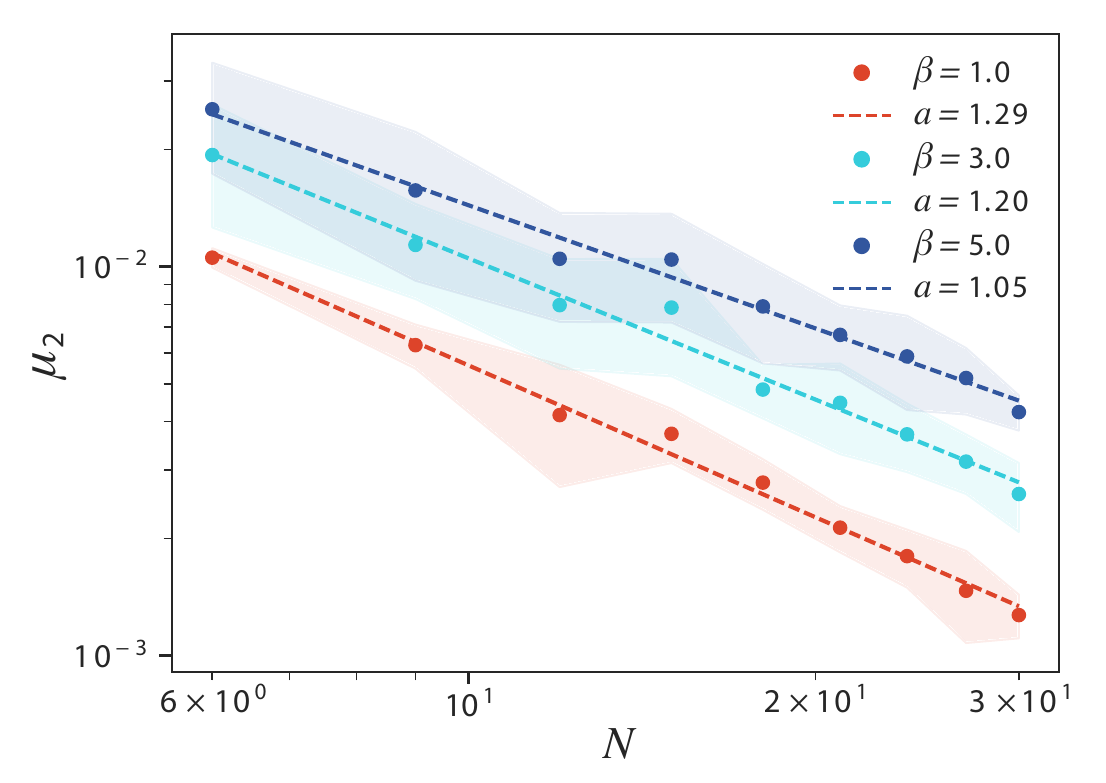}
        \caption{The effect of the constraints on the algebraic connectivity for different budget multipliers, $\beta$. We show the average and standard deviation of the algebraic connectivity for the optimized sampled graphs under different budget relaxations after $10$ independent runs. We assigned the number of agents to each sampled hypergraph to be four times the number of tasks in the hypergraph. The behavior of the algebraic connectivity follows $\mu_2 \sim N^{-a}$, where $N$ is the hypergraph size and $a$ is the scaling parameter.}
    \label{fig:Budget}
\end{figure}

We conducted an experiment to investigate the effect of constraints on algebraic connectivity. In this study, we analyzed the optimization results for the APS 2020-2021 dataset under relaxed agent budget constraints. These constraints reflect real-world constraints on agent workloads. To introduce varying degrees of budget relaxation into the constrained optimization problem, we multiplied the original agent budget limits in the APS data by a series of budget multipliers, $\beta = 1, 3, 5$.
We randomly sampled sub-hypergraphs of different sizes for each budget multiplier, multiplied the agent's budget constraint by the budget multiplier, and optimized them using the greedy algorithm. In Fig.~\ref{fig:Budget}, we report the average and standard deviation of the algebraic connectivity for the optimized sampled graphs under different budget relaxations after $10$ independent runs. In particular, we assigned the number of agents to each sampled hypergraph four times the number of tasks in the hypergraph.

Our observations indicate that, on average, more relaxed budget constraints, which allow for a larger allocation of agents' working time, lead to higher algebraic connectivity. We noticed that in this experiment, $\mu_2 \sim N^{-1}$, and that as we increase the budget multipliers, the algebraic connectivity also tends to increase. This suggests that we can interpret algebraic connectivity and constraints as two opposing forces. In other words, by controlling the constraints, we observe that the higher the algebraic connectivity, the denser the hypergraph. So, by enforcing the constraints, we limit the space of possible hypergraphs and, thus, the algebraic connectivity.

\subsection{Results}
\label{sec:results}

\subsubsection{Optimizing the Algebraic Connectivity}

\begin{figure*}[t!]
    \centering
\includegraphics[width=\textwidth]{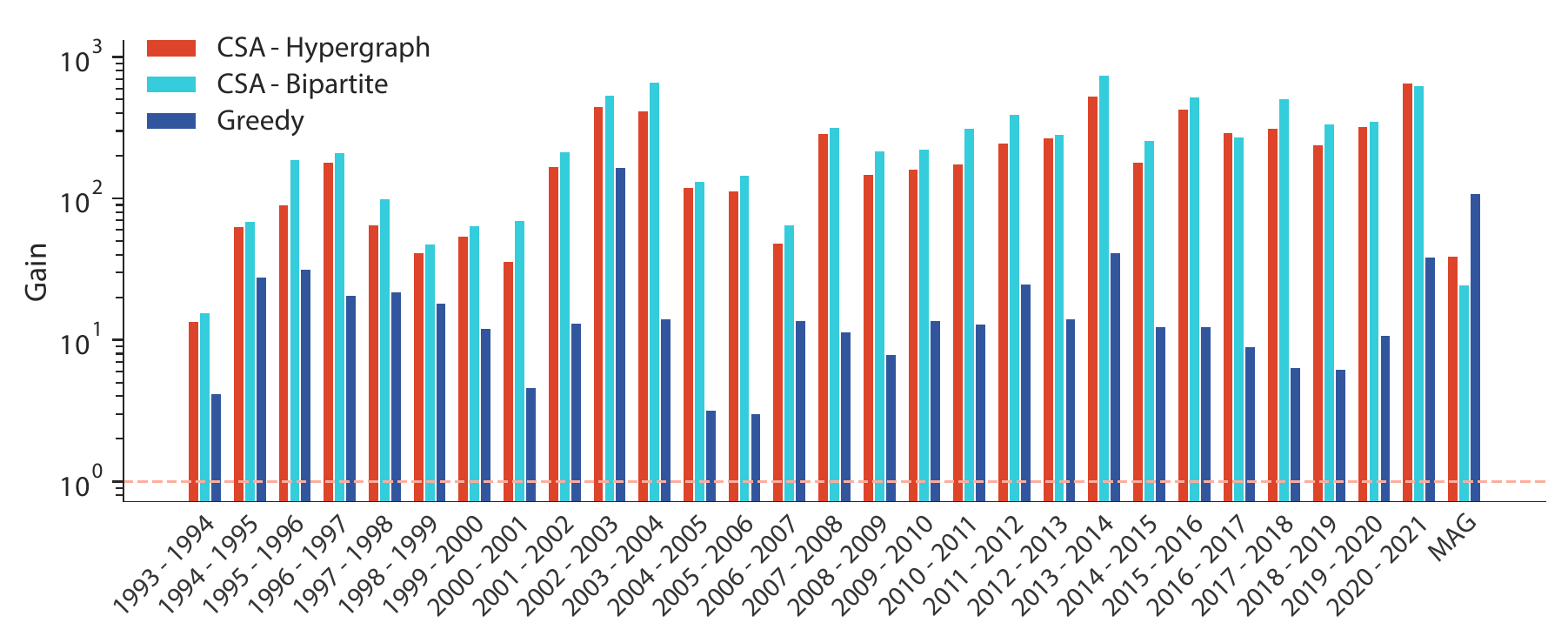}
    \caption{Algebraic connectivity gain for the extracted hypergraphs from the APS dataset between the years 1993-2021 and the MAG dataset, considering the original dataset, the CSA for the hypergraph and bipartite formulations, and the greedy approach.}
\label{fig:GainAPS}
\end{figure*}

\begin{figure*}[t!]
    \centering
\includegraphics[width=\textwidth]{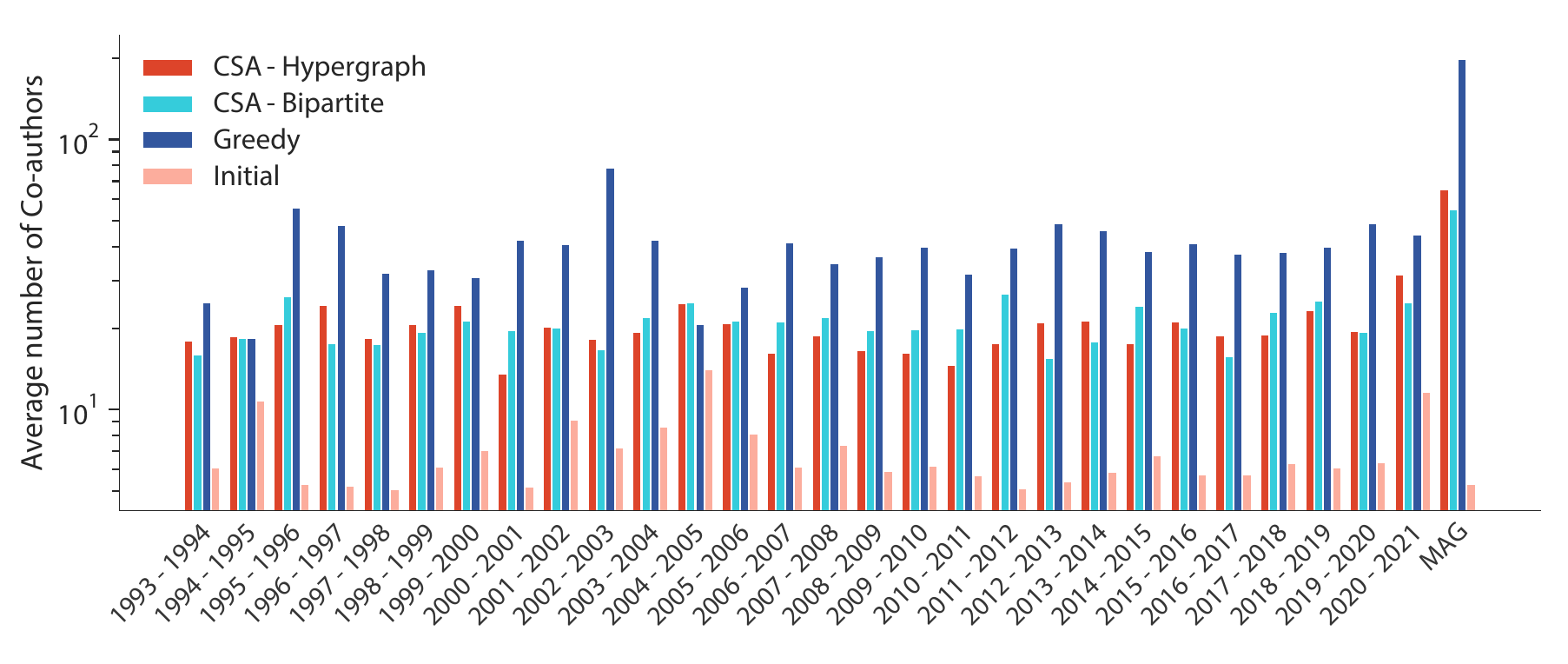}
    \caption{Average number of co-authors for the extracted hypergraphs from the APS dataset between the years 1993-2021 and the MAG dataset, considering the original dataset, the CSA for the hypergraph and bipartite formulations, and the greedy approach.}
\label{fig:AverageAuthorPaper}
\end{figure*}

\begin{figure*}[t!]
    \centering
\includegraphics[width=\textwidth]{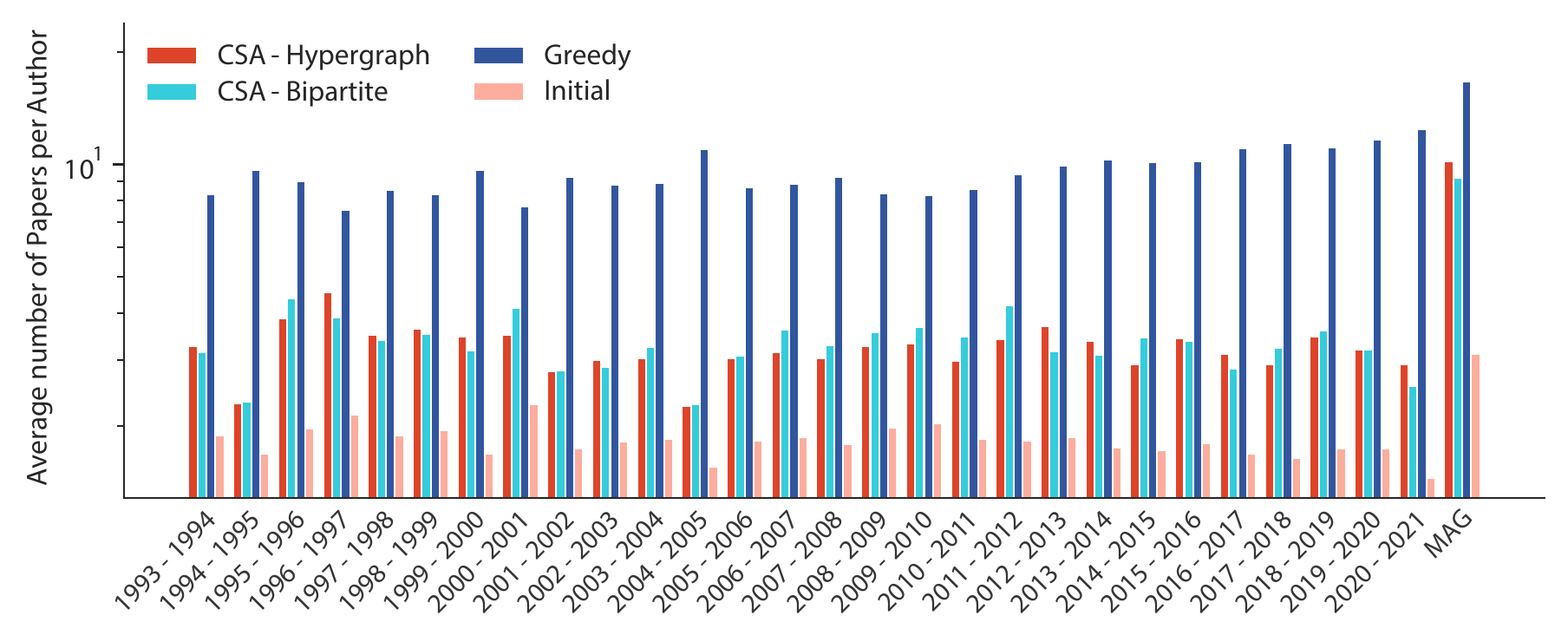}
    \caption{Average number of papers per author of the extracted hypergraphs from the APS dataset between the years 1993-2021 and the MAG dataset, considering the original dataset, the CSA for the hypergraph and bipartite formulations, and the greedy approach.}
\label{fig:AveragePaperAuthor}
\end{figure*}

In this section, we show our results on how we managed to optimize the algebraic connectivity of the hypergraphs extracted from real datasets. In Fig.~\ref{fig:GainAPS}, we present the algebraic connectivity gain, $Gain = \frac{\mu_2^{Optimized}}{\mu_2^{Real}}$, for the three optimization methods considered, the CSA for the hypergraph and bipartite formulations, and the greedy approach. Here, we consider the APS dataset with different extracted hypergraphs associated with 2-year periods from 1993 to 2021. Note that all approaches optimize the algebraic connectivity of the real collaboration hypergraph, as shown by the dashed line in Fig.~\ref{fig:GainAPS}. Notably, the CSA algorithm also provides a significantly better solution when compared to the greedy approach, where the algebraic connectivity in the optimized hypergraphs is between ten and five hundred times higher than the original hypergraph. In addition, in Fig.~\ref{fig:GainAPS}, we also show the comparison between the CSA approach using the hypergraph and the bipartite representation. Although they show approximately similar gains, we note that the comparison between the two methods is not straightforward because they represent different objects. The purpose of reporting such a comparison is to show that both systems are optimized versions of the original data. Thus, the comparisons are fair in terms of our quality measures.

We note that when applied to the 2002-2003 hypergraph, the greedy algorithm did not finish within the 500-hour time limit. To improve computational efficiency, we have developed an adaptation of the greedy algorithm. The adapted approach uses random assignment of tasks to agents when the number of available agents exceeds 50 during the assignment process. The rationale behind this adaptation is that early-stage assignments have a relatively small impact on the final optimized algebraic connectivity. This small adjustment sufficiently reduces the computational cost.

As mentioned in Section~\ref{sec:OPT}, we are also interested in some other quantities of our solutions, including the average number of tasks assigned to an agent, $\bar{T}$, and the average number of co-authors, $\hat{A}$. For the APS dataset, these two quantities are shown in the figures~\ref{fig:AverageAuthorPaper} and~\ref{fig:AveragePaperAuthor}, respectively. We notice that the optimized versions are systematically denser, both in terms of the average number of co-authors and the average number of papers per author (tasks). Interestingly, the CSA method for the hypergraph also provides solutions with a lower number of co-authors and a lower average number of papers per author compared to the greedy approach. This is a desirable feature, as we expect it to reduce communication costs.

\subsubsection{Resilience Against Attacks}

\begin{figure*}[t!]
    \centering
\includegraphics[width=\textwidth]{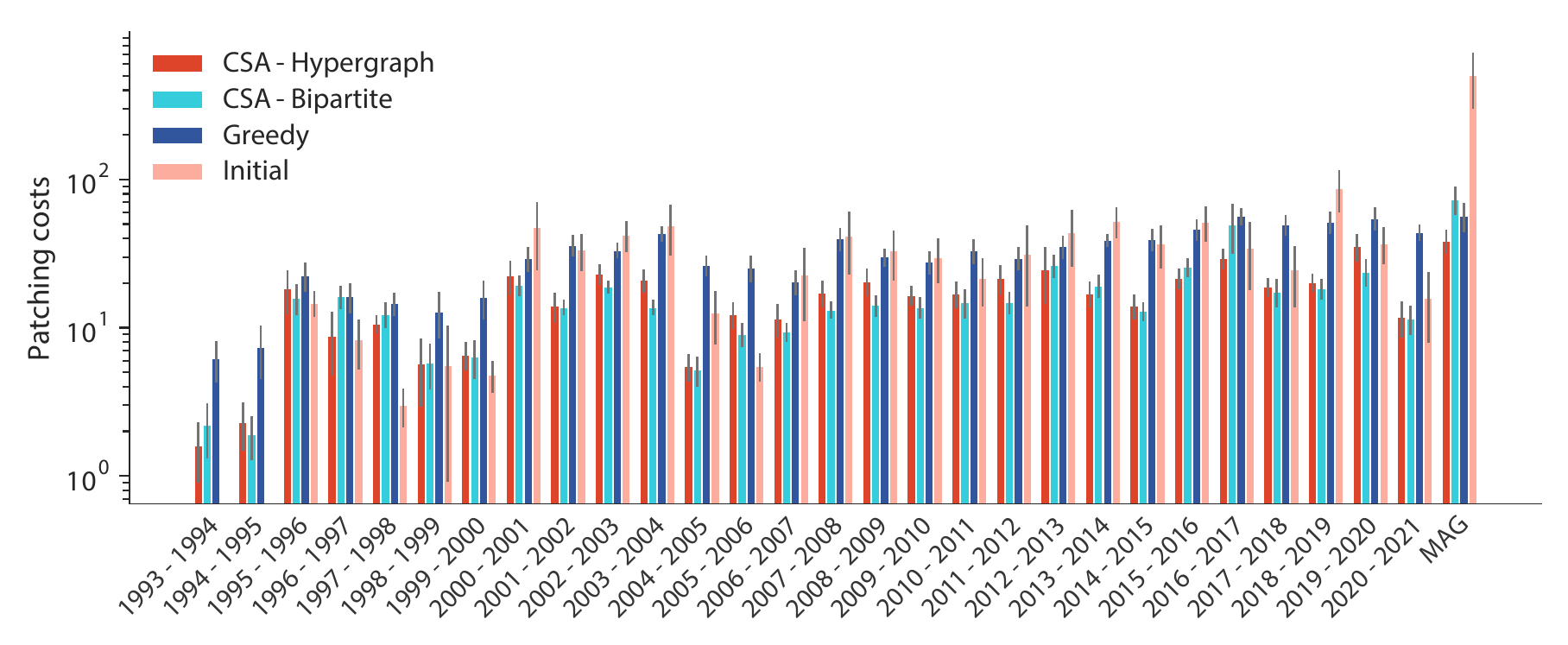}
    \caption{Patching costs after removing four nodes for the extracted hypergraphs from the APS dataset between the years 1993-2021 and the MAG dataset, considering the original dataset, the CSA for the hypergraph and bipartite formulations, and the greedy approach. The bars represent the average of $n_{exp} = 10$ runs, while the error bars represent the $\frac{\sigma_{exp}}{\sqrt{n_{exp}}}$. The patching costs are zero in the APS dataset for the hypergraphs of 1993-1994 and 1994-1995.}
    \centering
\label{fig:PatchingCosts4APS}
\end{figure*}

\begin{figure*}[t!]
    \centering
\includegraphics[width=\textwidth]{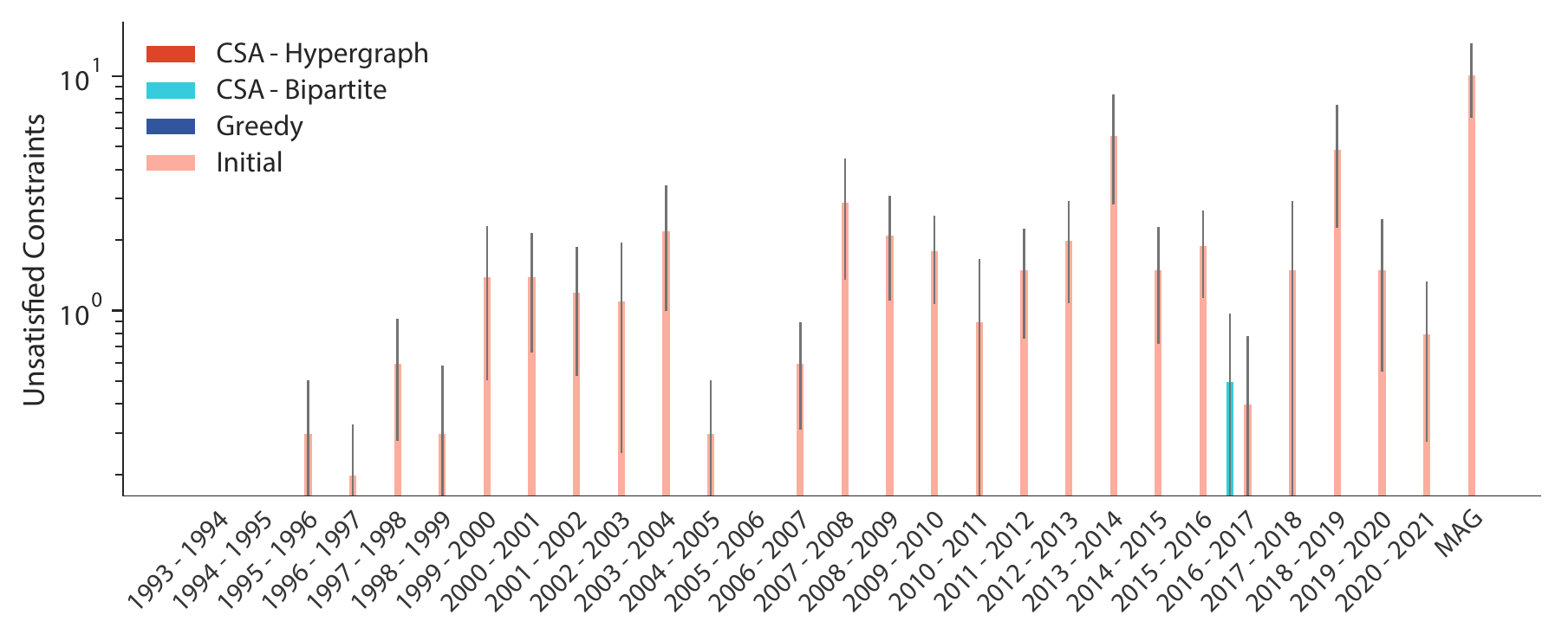}
    \caption{The sum of unsatisfied constraints after removing four nodes  and patching the solution  for the extracted hypergraphs from the APS dataset between the years 1993-2021 and the MAG dataset, considering the original dataset, the CSA for the hypergraph and bipartite formulations, and the greedy approach. The bars represent the average of $n_{exp} = 10$ runs, while the error bars represent the $\frac{\sigma_{exp}}{\sqrt{n_{exp}}}$. The unsatisfied constraints are zero for CSA hypergraph and bipartite (except for 2016-2017) and greedy for all of the cases (no bars are shown). }
\label{fig:UnsatCosts4APS}
\end{figure*}

In Fig.~\ref{fig:PatchingCosts4APS}, we show the patching cost of the optimized solution compared with the initial hypergraph under the four-node removal attack. The plots represent the patching costs of the hypergraphs corresponding to the APS dataset from 1993 to 2021. As we can see, the optimized solutions are more resilient to such attacks since the patching cost is almost always lower for the optimized solution. Similarly, we almost never have an unsuccessful patch for the optimized hypergraphs, while this is not the case for the original hypergraphs (results not shown). For comparison, we note that the greedy approach generally has a higher patching cost than the CSA approach. Also, both the hypergraph and bipartite approaches using the CSA method gave similar results.

In Fig.~\ref{fig:UnsatCosts4APS}, we show our main result, the sum of the unsatisfied constraints under the removal of four nodes and after patching the solution. We can see that in the CSA approach for the hypergraph, the removal of four nodes does not imply any violation of the constraints after patching. For the bipartite CSA case, the results are similar. The only exception is the hypergraph extracted from 2016-2017. However, this is not statistically sufficient to conclude that one approach is better than the other. Furthermore, we notice that the hypergraphs from 1993-1994 and 1994-1995 have zero unsatisfied constraints in the original data. This also explains why the original hypergraph has a lower patching cost than the optimized version. A similar analysis can be done for the 2005-2006 hypergraph. Moreover, we notice that the greedy approach seems to perform well as it does not leave unsatisfied constraints.

Similarly, in the appendix, we also provide the same results for the removal of two nodes, where we have similar results.

Finally, we report a similar analysis for the MAG dataset. Although it is difficult to directly compare these two datasets, they are similar in nature, and their main difference lies in how the data are collected, curated, and selected. For the latter, in the APS experiments, we fixed a 2-year time window and observed only the changes that occurred due to social factors, the number of researchers working in statistical mechanics, and their productivity. On the other hand, for the MAG dataset, the data were filtered using the keyword ``hypergraph'', and time is not a constraint. This factor may impose different constraints, as time is closely related to our notion of how many papers a researcher can produce. Note that this is translated in our models as our constraints, i.e., the energy agents can spend and the energy tasks that must be completed. Despite these differences, there are no methodological issues, and the comparisons are still reasonable due to their similar nature.

Regarding the MAG dataset (see the last set of bars in figures~\ref{fig:GainAPS},~\ref{fig:AverageAuthorPaper},~\ref{fig:AveragePaperAuthor},~\ref{fig:PatchingCosts4APS}, and~\ref{fig:UnsatCosts4APS}), the results are similar to those obtained for the APS, where patching costs and unsuccessful patches are significantly reduced. In the case of the MAG collaboration dataset, the algebraic connectivity gain is $ Gain = \frac{\mu_2^{Optimized}}{\mu_2^{Real}} \approx 39$, which is similar to the $Gain$ observed in the APS hypergraphs, as shown in Fig.~\ref{fig:GainAPS}. The same can be said for the average number of co-authors and the average number of papers per author (see figures~\ref{fig:AverageAuthorPaper} and~\ref{fig:AveragePaperAuthor}). The most important results concern patching costs and unsuccessful patches after the attack. We observe a significant reduction in the patching cost of the optimized case, about $60 \%$ lower on average.
However, we see a notable reduction in the number of unsuccessful patches and unsatisfied constraints since the numbers of unsuccessful patches and unsatisfied constraints were close to zero most of the time  for the optimized solutions, as can be seen in Fig.~\ref{fig:UnsatCosts4APS} and ~\ref{fig:UnsatCosts2APS}.
In general, the results on both datasets are in agreement, suggesting that our methods and results are robust to different datasets.

\section{Discussion}

\subsection{Algebraic Connectivity and the Team Assignment Problem}

Our motivation to propose the use of algebraic connectivity in the team assignment problem arises from its applications in graph theory~\cite{Jamakovic2008, Cozzo2019}. Thus, our main results concern the robustness of our assignment with respect to the patching costs and unsatisfied constraints, shown in figures~\ref{fig:PatchingCosts4APS} and~\ref{fig:UnsatCosts4APS}, respectively. From this analysis, we can see that the algebraic connectivity is indeed capturing the resilience features of the assignment. This is specifically evident when analyzing the unsatisfied constraints, Fig.~\ref{fig:UnsatCosts4APS}, where the CSA approaches have no unsatisfied constraints after the four-node attack and patching the solution.

Moreover, by optimizing the algebraic connectivity, we expect also to reduce the timescale of diffusion processes, as argued in Sec.~\ref{sec:Laplacian}. Although diffusion is just a mathematical model, in practice, we hope that such quantity can also be reflected in practical terms of information diffusion.
Our main concern in this case was the average number of agents per task and the average number of co-authors. As shown in figures~\ref{fig:AverageAuthorPaper} and~\ref{fig:AveragePaperAuthor}, these two measures tend to increase as a consequence of optimizing the algebraic connectivity. This can be observed by noticing that in all the tested optimization methods, these quantities increased. However, our CSA hypergraph approach presented an increase that is lower than that of the greedy method. Thus, our methods are improving the robustness and diffusion at the same time but without unboundedly increasing the communication costs.

We should remark that perhaps the algebraic connectivity alone might not be enough. Here, the constraints play a major role in our results. They reduce the space of feasible solutions and act as an opposing force, driving the solutions towards more practical solutions. We also remark that setting appropriate solutions must be key in real applications.

\subsection{Comparison with the Baseline Greedy Solutions}

Comparing the greedy with the original data, we notice an improvement in algebraic connectivity, Fig.~\ref{fig:GainAPS}.
However, this comes at the cost of higher values for the average number of agents per task and the average number of co-authors; see figures~\ref{fig:AverageAuthorPaper} and~\ref{fig:AveragePaperAuthor}.
When analyzing the patching costs, we observe inconsistent results of both an improvement and degradation of the solutions in terms of patching costs after the four-node attack experiment.
However, similar to CSA, the greedy approach presents satisfying results in terms of unsatisfied constraints after node removal attacks and patching, see Fig.~\ref{fig:UnsatCosts4APS}.

From a computational point of view, the greedy approach is $O(N^4)$, while the computation of the algebraic connectivity alone is $O(N^3)$.
Due to the stochastic nature of the CSA approaches, the comparisons may be perceived as unfair since one could run the CSA for an arbitrary number of iterations. However, the overall quality of the greedy solutions was not satisfactory. Finally, one could use mixed approaches where the greedy solution is used as the initial state for the CSA.

\subsection{Comparison with the Bipartite Representation}

The CSA approaches for the hypergraph and the bipartite formulations seem to be statistically equivalent in our experiments, both in terms of algebraic connectivity gain, Fig.~\ref{fig:GainAPS}, and robustness, Figs~\ref{fig:PatchingCosts4APS} and~\ref{fig:UnsatCosts4APS}. Thus, we have no evidence to advocate one approach over the other in terms of assignments. However, the computational cost of computing the algebraic connectivity for the bipartite approach is significantly higher, $O((N+K)^3)$, versus $O(N^3)$ in the hypergraph case. Thus, we have evidence that the hypergraph approach should be preferred in practice. Note that with the same computational resources, we should be closer to the global optima in the hypergraph case than in the bipartite case since we should be able to explore a larger space.

\subsection{Hyperparameter Selection}

There are several hyperparameters in Alg.~\ref{alg:CSA}, and in this section, we will explain the sensitivity of our algorithm to them and how one should choose their values. Some of the hyperparameters, such as the initial temperature ($T_0$) and the cooling schedule factor ($a_c$), are related to simulated annealing. These two hyperparameters can be chosen similarly to any other simulated annealing algorithm. One must be careful not to choose very small values for them to allow exploration. On the other hand, very large values for these two parameters will cause the algorithm to produce subpar solutions. We also have two other hyperparameters, $t_{max}$ and $T_{th}$, which determine when the algorithm should stop. It is easy to change these two parameters to allow the algorithm to converge to a good solution before stopping.

We also have some hyperparameters that act as penalty coefficients for the penalty functions that we add. We add two types of penalty functions to our objective function. The first penalty function is to penalize the infeasible solutions. The second penalty function is to control the two important factors of $\bar{T}$ (average number of tasks per agent), and $\hat{A}$ (the average number of teammates per agent) during the optimization. The first penalty function does not play a major role in the optimization process because of the guided perturbation approach used. Regarding the second penalty function, with a higher coefficient, we get a smaller final algebraic connectivity, but also smaller $\bar{T}$ and $\hat{A}$. One can adjust this coefficient depending on how much one can tolerate large values of these factors. 


\subsection{Alternative Optimization Methods}

Although we have used Constrained Simulated Annealing (Alg.~\ref{alg:CSA}) as the main optimization method in this paper, we have also considered and tried (where possible) other approaches. The black-box nature of our objective function (we only have access to zero-order information) prevents us from using optimization algorithms that require a closed-form expression or the first-order gradient information of the objective function. Gradient descent-based algorithms (SGD, ADAM, etc.)~\cite{kingma2014adam} require the first-order gradient information of the objective function, and therefore we cannot use them for our problem. There are also some recent unsupervised learning-based approaches for solving the optimization problems~\cite{schuetz2022combinatorial, heydaribeni2023hypop}, but they still need the gradient information of the objective. The supervised learning-based approaches were also not feasible for our problem because they require a training dataset of solved problem instances~\cite{cappart2023combinatorial}.

We also looked at other optimization tools such as Gurobi optimization~\cite{Gurobi}. However, since Gurobi only works with objective functions that are either linear, piecewise linear, or quadratic, we could not use it for our problem.

We had two other viable alternatives. The first alternative was to use a greedy algorithm, which we used as the basis for our experiments. The second alternative was to use a sampling-based black-box optimization method such as AdaNS~\cite{javaheripi2020adans}, and we tried to solve our optimization problem with this method. Similar to what we did in Alg.~\ref{alg:CSA}, we added penalty functions to enforce feasible solutions. However, we noticed that the ratio of the feasible region to the entire search space is considerably small, and consequently, the sampling-based approach failed to find the feasible region. We realized that the samples taken in such algorithms must be directed towards the feasible region; otherwise, they will get lost in the infeasible parts of the search space. The simulated annealing algorithm provided a better and simpler basis for us to modify the sampling phase, so we decided to use simulated annealing with a guided sampling algorithm that we designed to solve our optimization problem.

\section{Conclusion}

We propose a team assignment algorithm based on a hypergraph approach that focuses on resilience and diffusion optimization.
More specifically, we map the effort of each agent in a task as an edge-dependent vertex-weighted hypergraph. Our approach is based on optimizing the algebraic connectivity of such a hypergraph.
In our formulation, we also consider two constraints: the energy expended by each agent and the energy required to complete a task.
These constraints reduce the feasible region and act as an opposing force to the algebraic connectivity. In practice, we used constrained simulated annealing to find the optimal solution.

We systematically evaluated all connected small hypergraphs with $N=5$ agents and $K=3$ tasks to validate the algebraic connectivity. This experiment showed that algebraic connectivity favors densely connected hypergraphs. In addition, we performed a finite-size analysis considering four classes of hypergraphs: (i) connected by a node, (ii) connected by a hyperedge, (iii) head-to-tail, and (iv) randomly swapped hypergraphs.
This analysis verifies that head-to-tail structures scale worse (larger exponent) than the centralized structures, i.e., centralized by nodes or hyperedge. In practice, these two features will drive the optimization algorithm towards a robust assignment. Note that they are complemented by the constraints in our model.

We tested our methods on two scientific collaboration datasets, the MAG and the APS. We evaluated the robustness of our assignment using an attack-based evaluation, where nodes are removed, and we estimated the cost of moving the assignment into the feasible region. We verified that our optimized hypergraphs are significantly more resilient than the original data. In addition, we compared our constraint simulated annealing approach with a greedy approach. In this case, the CSA shows a significant improvement over the greedy approach in terms of algebraic connectivity.

We have also compared our hypergraph approach with a bipartite version that captures similar properties. Namely, the random walk defined on the hypergraph is the same as a two-step random walk in the bipartite. We verified that both approaches give similar results in terms of attacks and unsuccessful patching costs. However, the computational cost of the bipartite approach is significantly higher, $O((N+K)^3)$, versus $O(N^3)$ in the hypergraph case.

We hope that our results motivate further exploration of algebraic connectivity in the team assignment problem. Our approach does not include a skill set for the agents, but this feature can be incorporated as an additional set of constraints and is left as future work. A similar argument can be made for any other personalized algorithm. Another possibility would be the design of multi-objective optimizations where supporting fitness functions can be used. For example, one could think about simultaneously optimizing algebraic connectivity and energy consumption.

In addition, we believe that our methods, and possibly variations of them, can be used to generate what-if scenarios, as suggested in~\cite{juarez_comprehensive_2021}. In this case, the algebraic connectivity and its optimizations can quantify the resilience of the assignments. Finally, the proposed hypergraph mapping and algebraic connectivity can be explored to analyze other systems.
As a concrete example, we can mention financial systems similar to~\cite{Huang2013, Caccioli2014}. We recall that a dataset that can be seen as a bipartite graph can easily be mapped as an EDVW hypergraph. Thus, we hope that our methods and results can be useful in the analysis of real hypergraphs and other practical scenarios.

\begin{acknowledgments}
We thank M. Clarin for help with the figures. We acknowledge the support of the AccelNet-MultiNet program, a project of the National Science Foundation (Award \#1927425 and \#1927418) and also ARO MURI program (award number W911NF-21-1- 0322). 
G.F.A acknowledges Henrique Ferraz de Arruda for fruitful discussions and his help with the collaboration datasets.
Y.M was partially supported by the Government of Arag\'on, Spain and ``ERDF A way of making Europe'' through grant E36‐23R (FENOL), and by Ministerio de Ciencia e Innovaci\'on, Agencia Espa\~nola de Investigaci\'on (MCIN/AEI/10.13039/501100011033) Grant No. PID2020‐115800GB‐I00. We acknowledge the use of the computational resources of COSNET Lab at Institute BIFI, funded by Banco Santander (grant Santander‐UZ 2020/0274) and by the Government of Arag\'on (grant UZ-164255).
The funders had no role in study design, data collection and analysis, the decision to publish, or the preparation of the manuscript.
\end{acknowledgments}

\appendix

\section{Complementary Experiment: 2-node Attack}
\label{sec:2node}

\begin{figure*}[t!]
    \centering
\includegraphics[width=\textwidth]{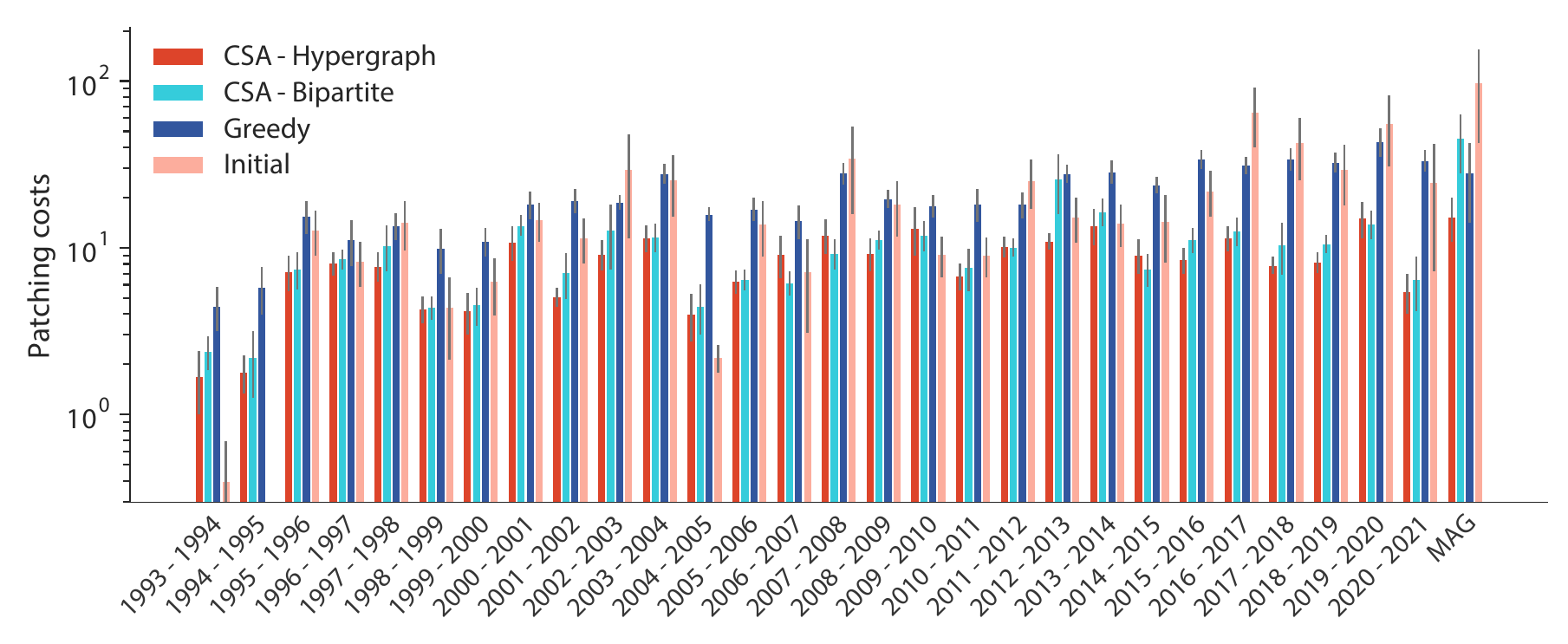}
    \caption{Patching costs after removing two nodes for the extracted hypergraphs from the APS dataset between the years 1993-2021, considering the original dataset, the CSA for the hypergraph and bipartite formulations, and the greedy approach. The bars represent the average of $n_{exp} = 10$ runs, while the error bars represent the $\frac{\sigma_{exp}}{\sqrt{n_{exp}}}$. The patching costs are zero in the APS dataset for the hypergraphs of 1994-1995.}
\label{fig:PatchingCosts2APS}
\end{figure*}

\begin{figure*}[t!]
    \centering
\includegraphics[width=\textwidth]{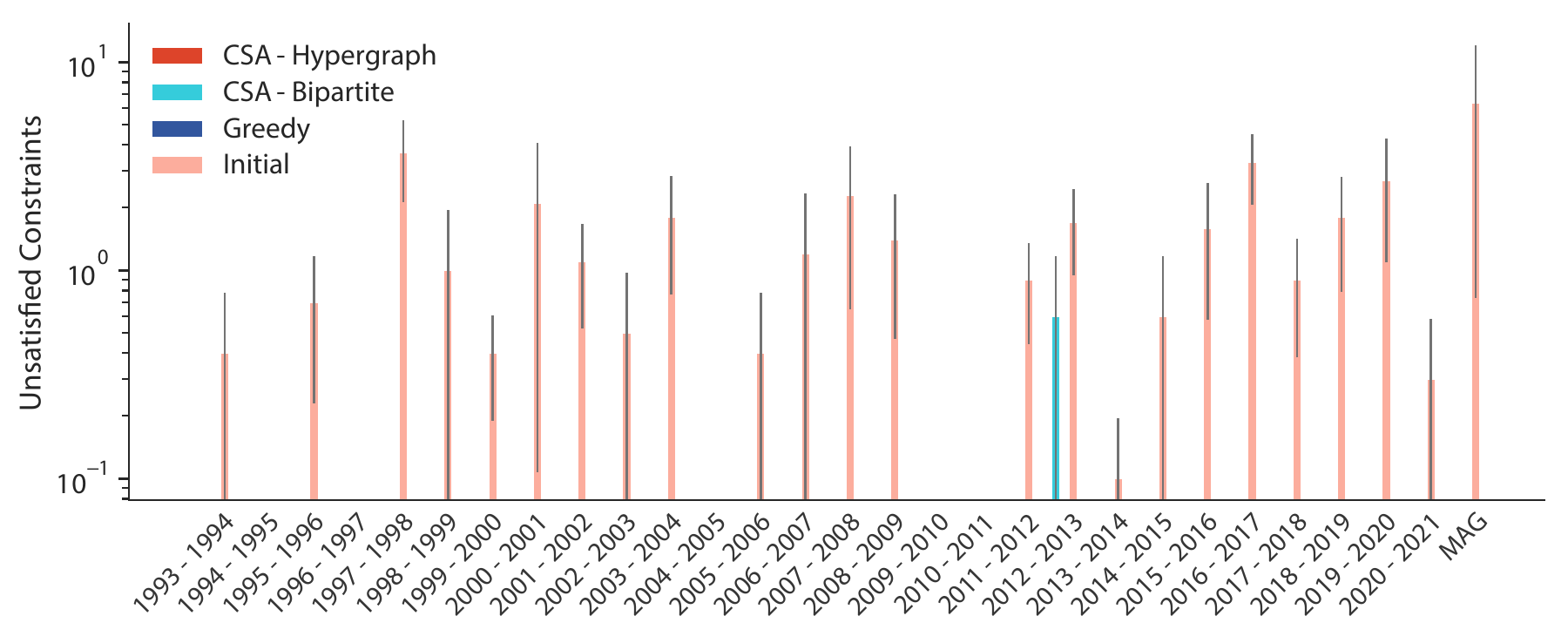}
    \caption{The sum of unsatisfied constraints after removing two nodes and patching the solution for the extracted hypergraphs from the APS dataset between the years 1993-2021, considering the original dataset, the CSA for the hypergraph and bipartite formulations, and the greedy approach. The bars represent the average of $n_{exp} = 10$ runs, while the error bars represent the $\frac{\sigma_{exp}}{\sqrt{n_{exp}}}$.  The unsatisfied constraints are zero for CSA hypergraph and bipartite (except for 2012-2013) and greedy for all of the cases (no bars are shown).}
\label{fig:UnsatCosts2APS}
\end{figure*}

In Figures~\ref{fig:PatchingCosts2APS}, and~\ref{fig:UnsatCosts2APS}, we show the patching costs and the unsatisfied constraints after removing two nodes and patching, respectively. Similar to the experiment reported in the main text, the CSA approach performs very well as it leaves no unsatisfied constraints. This observation is true for both the hypergraph and bipartite approaches. The only exception is the case of 2012-2013, where the bipartite approach leaves some unsatisfied constraints.
Since we observed a similar effect in the four-node attack, one might be tempted to say that the bipartite approach is less robust than the hypergraph case. However, these experiments may not be sufficient to make such a claim. On the other hand, the greedy approach produces a poor result, often requiring more patching costs than the initial data as opposed to both CSA  hypergraph and bipartite formulations that outperform the initial data and the greedy approach in terms of the patching cost.

\section{Individual hypergraphs description}
Table~\ref{tab:description} describes the hypergraphs extracted from the APS and MAG collaboration datasets. In this table, we have summarized the number of authors and papers, as well as the average budget and ``energy'' spent. The comparison between different hypergraphs is not straightforward, as we should consider a number of external and internal factors. As external factors, we should mention that although the APS hypergraphs have a fixed time window, the interest and productivity of researchers in that period can vary depending on many factors. As internal factors, we have the keyword used to extract the MAG hypergraph and the fact that we use only the giant connected component. Note that for the MAG, this comparison is even more difficult since the fields may only have a small overlap (the keyword hypergraph in the MAG vs. statistical mechanics published in PRE).

\begin{table}[!ht]
    \centering
    \begin{tabular}{|c|c|c|c|c|}
        \hline
        Hypergraph & $N$ & $K$ & $ \mathbb{E}[ B ]$ & $\mathbb{E}[ E ]$ \\
        \hline
        \hline
        APS 1993 - 1994 & $52$ & $25$ & $7.200$ & $3.462$ \\ \hline
        APS 1994 - 1995 & $92$ & $25$ & $7.320$ & $1.989$ \\ \hline
        APS 1995 - 1996 & $241$ & $153$ & $8.163$ & $5.183$ \\ \hline
        APS 1996 - 1997 & $168$ & $122$ & $6.738$ & $4.893$ \\ \hline
        APS 1997 - 1998 & $148$ & $86$ & $7.698$ & $4.473$ \\ \hline
        APS 1998 - 1999 & $107$ & $58$ & $7.328$ & $3.972$ \\ \hline
        APS 1999 - 2000 & $163$ & $69$ & $8.304$ & $3.515$ \\ \hline
        APS 2000 - 2001 & $312$ & $235$ & $6.826$ & $5.141$ \\ \hline
        APS 2001 - 2002 & $553$ & $221$ & $7.484$ & $2.991$ \\ \hline
        APS 2002 - 2003 & $1344$ & $650$ & $7.561$ & $3.657$ \\ \hline
        APS 2003 - 2004 & $991$ & $471$ & $7.411$ & $3.522$ \\ \hline
        APS 2004 - 2005 & $369$ & $82$ & $7.695$ & $1.710$ \\ \hline
        APS 2005 - 2006 & $313$ & $133$ & $7.150$ & $3.038$ \\ \hline
        APS 2006 - 2007 & $216$ & $113$ & $7.602$ & $3.977$ \\ \hline
        APS 2007 - 2008 & $707$ & $330$ & $7.848$ & $3.663$ \\ \hline
        APS 2008 - 2009 & $570$ & $330$ & $7.173$ & $4.153$ \\ \hline
        APS 2009 - 2010 & $370$ & $220$ & $7.236$ & $4.303$ \\ \hline
        APS 2010 - 2011 & $376$ & $201$ & $7.368$ & $3.939$ \\ \hline
        APS 2011 - 2012 & $374$ & $218$ & $8.351$ & $4.868$ \\ \hline
        APS 2012 - 2013 & $538$ & $310$ & $9.003$ & $5.188$ \\ \hline
        APS 2013 - 2014 & $647$ & $317$ & $8.991$ & $4.405$ \\ \hline
        APS 2014 - 2015 & $559$ & $250$ & $8.820$ & $3.945$ \\ \hline
        APS 2015 - 2016 & $752$ & $386$ & $9.049$ & $4.645$ \\ \hline
        APS 2016 - 2017 & $962$ & $456$ & $9.774$ & $4.633$ \\ \hline
        APS 2017 - 2018 & $825$ & $347$ & $9.968$ & $4.193$ \\ \hline
        APS 2018 - 2019 & $895$ & $424$ & $9.837$ & $4.660$ \\ \hline
        APS 2019 - 2020 & $1162$ & $569$ & $10.501$ & $5.142$ \\ \hline
        APS 2020 - 2021 & $769$ & $201$ & $10.328$ & $2.700$ \\ \hline
        MAG    & $781$ & $704$ & $16.672$ &  $15.028$  \\ \hline
    \end{tabular}
    \caption{Summary of the hypergraphs extracted from the APS and MAG collaboration datasets. $N$ = number of authors (a.k.a.~agents). $K$ = number of papers (a.k.a.~tasks). $ \mathbb{E}[ B ]$ = average budget (average number of papers per author) 
and $\mathbb{E}[ E ]$ = average energy (average number of authors per paper). }
    \label{tab:description}
\end{table}

\bibliographystyle{apsrev}
\bibliography{main}

\begin{thebibliography}{53}
\expandafter\ifx\csname natexlab\endcsname\relax\def\natexlab#1{#1}\fi
\expandafter\ifx\csname bibnamefont\endcsname\relax
  \def\bibnamefont#1{#1}\fi
\expandafter\ifx\csname bibfnamefont\endcsname\relax
  \def\bibfnamefont#1{#1}\fi
\expandafter\ifx\csname citenamefont\endcsname\relax
  \def\citenamefont#1{#1}\fi
\expandafter\ifx\csname url\endcsname\relax
  \def\url#1{\texttt{#1}}\fi
\expandafter\ifx\csname urlprefix\endcsname\relax\def\urlprefix{URL }\fi
\providecommand{\bibinfo}[2]{#2}
\providecommand{\eprint}[2][]{\url{#2}}

\bibitem[{\citenamefont{Juárez et~al.}(2021)\citenamefont{Juárez, Santos, and Brizuela}}]{juarez_comprehensive_2021}
\bibinfo{author}{\bibfnamefont{J.}~\bibnamefont{Juárez}}, \bibinfo{author}{\bibfnamefont{C.~P.} \bibnamefont{Santos}}, \bibnamefont{and} \bibinfo{author}{\bibfnamefont{C.~A.} \bibnamefont{Brizuela}}, \bibinfo{journal}{ACM Computing Surveys} \textbf{\bibinfo{volume}{54}}, \bibinfo{pages}{153:1} (\bibinfo{year}{2021}), ISSN \bibinfo{issn}{0360-0300}.

\bibitem[{\citenamefont{Kuhn}(1955)}]{Kuhn1955Hungarian}
\bibinfo{author}{\bibfnamefont{H.~W.} \bibnamefont{Kuhn}}, \bibinfo{journal}{Naval Research Logistics Quarterly} \textbf{\bibinfo{volume}{2}}, \bibinfo{pages}{83} (\bibinfo{year}{1955}).

\bibitem[{\citenamefont{Ahmed et~al.}(2013)\citenamefont{Ahmed, Deb, and Jindal}}]{Abhilash2013}
\bibinfo{author}{\bibfnamefont{F.}~\bibnamefont{Ahmed}}, \bibinfo{author}{\bibfnamefont{K.}~\bibnamefont{Deb}}, \bibnamefont{and} \bibinfo{author}{\bibfnamefont{A.}~\bibnamefont{Jindal}}, \bibinfo{journal}{Appl. Soft Comput.} \textbf{\bibinfo{volume}{13}}, \bibinfo{pages}{402–414} (\bibinfo{year}{2013}), ISSN \bibinfo{issn}{1568-4946}.

\bibitem[{\citenamefont{Okimoto et~al.}(2015)\citenamefont{Okimoto, Schwind, Clement, Ribeiro, Inoue, and Marquis}}]{Okimoto2015}
\bibinfo{author}{\bibfnamefont{T.}~\bibnamefont{Okimoto}}, \bibinfo{author}{\bibfnamefont{N.}~\bibnamefont{Schwind}}, \bibinfo{author}{\bibfnamefont{M.}~\bibnamefont{Clement}}, \bibinfo{author}{\bibfnamefont{T.}~\bibnamefont{Ribeiro}}, \bibinfo{author}{\bibfnamefont{K.}~\bibnamefont{Inoue}}, \bibnamefont{and} \bibinfo{author}{\bibfnamefont{P.}~\bibnamefont{Marquis}}, in \emph{\bibinfo{booktitle}{Proceedings of the 2015 International Conference on Autonomous Agents and Multiagent Systems}} (\bibinfo{publisher}{International Foundation for Autonomous Agents and Multiagent Systems}, \bibinfo{address}{Richland, SC}, \bibinfo{year}{2015}), AAMAS '15, p. \bibinfo{pages}{395–403}.

\bibitem[{\citenamefont{Okimoto et~al.}(2016)\citenamefont{Okimoto, Ribeiro, Bouchabou, and Inoue}}]{Okimoto2016M}
\bibinfo{author}{\bibfnamefont{T.}~\bibnamefont{Okimoto}}, \bibinfo{author}{\bibfnamefont{T.}~\bibnamefont{Ribeiro}}, \bibinfo{author}{\bibfnamefont{D.}~\bibnamefont{Bouchabou}}, \bibnamefont{and} \bibinfo{author}{\bibfnamefont{K.}~\bibnamefont{Inoue}}, in \emph{\bibinfo{booktitle}{International Joint Conference on Artificial Intelligence}} (\bibinfo{year}{2016}).

\bibitem[{\citenamefont{Demirovi\'{c} et~al.}(2018)\citenamefont{Demirovi\'{c}, Schwind, Okimoto, and Inoue}}]{Katsumi2018}
\bibinfo{author}{\bibfnamefont{E.}~\bibnamefont{Demirovi\'{c}}}, \bibinfo{author}{\bibfnamefont{N.}~\bibnamefont{Schwind}}, \bibinfo{author}{\bibfnamefont{T.}~\bibnamefont{Okimoto}}, \bibnamefont{and} \bibinfo{author}{\bibfnamefont{K.}~\bibnamefont{Inoue}}, in \emph{\bibinfo{booktitle}{Proceedings of the 17th International Conference on Autonomous Agents and MultiAgent Systems}} (\bibinfo{publisher}{International Foundation for Autonomous Agents and Multiagent Systems}, \bibinfo{address}{Richland, SC}, \bibinfo{year}{2018}), AAMAS '18, p. \bibinfo{pages}{1362–1370}.

\bibitem[{\citenamefont{Kargar et~al.}(2023)\citenamefont{Kargar, Zihayat, and An}}]{Aijun2013}
\bibinfo{author}{\bibfnamefont{M.}~\bibnamefont{Kargar}}, \bibinfo{author}{\bibfnamefont{M.}~\bibnamefont{Zihayat}}, \bibnamefont{and} \bibinfo{author}{\bibfnamefont{A.}~\bibnamefont{An}}, \emph{\bibinfo{title}{Finding Affordable and Collaborative Teams from a Network of Experts}} (\bibinfo{year}{2023}), pp. \bibinfo{pages}{587--595}.

\bibitem[{\citenamefont{Golshan et~al.}(2014)\citenamefont{Golshan, Lappas, and Terzi}}]{Evimaria2014}
\bibinfo{author}{\bibfnamefont{B.}~\bibnamefont{Golshan}}, \bibinfo{author}{\bibfnamefont{T.}~\bibnamefont{Lappas}}, \bibnamefont{and} \bibinfo{author}{\bibfnamefont{E.}~\bibnamefont{Terzi}}, in \emph{\bibinfo{booktitle}{Proceedings of the 20th ACM SIGKDD International Conference on Knowledge Discovery and Data Mining}} (\bibinfo{publisher}{Association for Computing Machinery}, \bibinfo{address}{New York, NY, USA}, \bibinfo{year}{2014}), KDD '14, p. \bibinfo{pages}{1196–1205}, ISBN \bibinfo{isbn}{9781450329569}.

\bibitem[{\citenamefont{Liu et~al.}(2015)\citenamefont{Liu, Luo, Tang, and Bressan}}]{Qing2015}
\bibinfo{author}{\bibfnamefont{Q.}~\bibnamefont{Liu}}, \bibinfo{author}{\bibfnamefont{T.}~\bibnamefont{Luo}}, \bibinfo{author}{\bibfnamefont{R.}~\bibnamefont{Tang}}, \bibnamefont{and} \bibinfo{author}{\bibfnamefont{S.}~\bibnamefont{Bressan}}, in \emph{\bibinfo{booktitle}{2015 IEEE International Conference on Communications (ICC)}} (\bibinfo{year}{2015}), pp. \bibinfo{pages}{567--572}.

\bibitem[{\citenamefont{ZAKARIAN and KUSIAK}(1999)}]{KUSIAK1999}
\bibinfo{author}{\bibfnamefont{A.}~\bibnamefont{ZAKARIAN}} \bibnamefont{and} \bibinfo{author}{\bibfnamefont{A.}~\bibnamefont{KUSIAK}}, \bibinfo{journal}{IIE Transactions} \textbf{\bibinfo{volume}{31}}, \bibinfo{pages}{85} (\bibinfo{year}{1999}).

\bibitem[{\citenamefont{Fitzpatrick and Askin}(2005)}]{Fitzpatrick2005}
\bibinfo{author}{\bibfnamefont{E.}~\bibnamefont{Fitzpatrick}} \bibnamefont{and} \bibinfo{author}{\bibfnamefont{R.~G.} \bibnamefont{Askin}}, \bibinfo{journal}{Comput. Ind. Eng.} \textbf{\bibinfo{volume}{48}}, \bibinfo{pages}{593} (\bibinfo{year}{2005}).

\bibitem[{\citenamefont{Feng et~al.}(2010)\citenamefont{Feng, Jiang, Fan, and Fu}}]{Feng2010}
\bibinfo{author}{\bibfnamefont{B.}~\bibnamefont{Feng}}, \bibinfo{author}{\bibfnamefont{Z.-Z.} \bibnamefont{Jiang}}, \bibinfo{author}{\bibfnamefont{Z.-P.} \bibnamefont{Fan}}, \bibnamefont{and} \bibinfo{author}{\bibfnamefont{N.}~\bibnamefont{Fu}}, \bibinfo{journal}{European Journal of Operational Research} \textbf{\bibinfo{volume}{203}}, \bibinfo{pages}{652} (\bibinfo{year}{2010}), ISSN \bibinfo{issn}{0377-2217}.

\bibitem[{\citenamefont{Chen and Lin}(2004)}]{Lin2004}
\bibinfo{author}{\bibfnamefont{S.-J.} \bibnamefont{Chen}} \bibnamefont{and} \bibinfo{author}{\bibfnamefont{L.}~\bibnamefont{Lin}}, \bibinfo{journal}{IEEE Transactions on Engineering Management} \textbf{\bibinfo{volume}{51}}, \bibinfo{pages}{111} (\bibinfo{year}{2004}).

\bibitem[{\citenamefont{Kargar and An}(2011)}]{Aijun2011}
\bibinfo{author}{\bibfnamefont{M.}~\bibnamefont{Kargar}} \bibnamefont{and} \bibinfo{author}{\bibfnamefont{A.}~\bibnamefont{An}}, in \emph{\bibinfo{booktitle}{Proceedings of the 20th ACM International Conference on Information and Knowledge Management}} (\bibinfo{publisher}{Association for Computing Machinery}, \bibinfo{address}{New York, NY, USA}, \bibinfo{year}{2011}), CIKM '11, p. \bibinfo{pages}{985–994}.

\bibitem[{\citenamefont{Lappas et~al.}(2009)\citenamefont{Lappas, Liu, and Terzi}}]{Lappas2009}
\bibinfo{author}{\bibfnamefont{T.}~\bibnamefont{Lappas}}, \bibinfo{author}{\bibfnamefont{K.}~\bibnamefont{Liu}}, \bibnamefont{and} \bibinfo{author}{\bibfnamefont{E.}~\bibnamefont{Terzi}}, in \emph{\bibinfo{booktitle}{Proceedings of the 15th ACM SIGKDD International Conference on Knowledge Discovery and Data Mining}} (\bibinfo{publisher}{Association for Computing Machinery}, \bibinfo{address}{New York, NY, USA}, \bibinfo{year}{2009}), KDD '09, p. \bibinfo{pages}{467–476}, ISBN \bibinfo{isbn}{9781605584959}.

\bibitem[{\citenamefont{Iacopini et~al.}(2019)\citenamefont{Iacopini, Petri, Barrat, and Latora}}]{Iacopo2019}
\bibinfo{author}{\bibfnamefont{I.}~\bibnamefont{Iacopini}}, \bibinfo{author}{\bibfnamefont{G.}~\bibnamefont{Petri}}, \bibinfo{author}{\bibfnamefont{A.}~\bibnamefont{Barrat}}, \bibnamefont{and} \bibinfo{author}{\bibfnamefont{V.}~\bibnamefont{Latora}}, \bibinfo{journal}{{Nature Communications}} \textbf{\bibinfo{volume}{10}}, \bibinfo{pages}{1} (\bibinfo{year}{2019}).

\bibitem[{\citenamefont{Jhun et~al.}(2019)\citenamefont{Jhun, Jo, and Kahng}}]{Jhun_2019}
\bibinfo{author}{\bibfnamefont{B.}~\bibnamefont{Jhun}}, \bibinfo{author}{\bibfnamefont{M.}~\bibnamefont{Jo}}, \bibnamefont{and} \bibinfo{author}{\bibfnamefont{B.}~\bibnamefont{Kahng}}, \bibinfo{journal}{Journal of Statistical Mechanics: Theory and Experiment} \textbf{\bibinfo{volume}{2019}}, \bibinfo{pages}{123207} (\bibinfo{year}{2019}).

\bibitem[{\citenamefont{de~Arruda et~al.}(2020)\citenamefont{de~Arruda, Petri, and Moreno}}]{Arruda2020}
\bibinfo{author}{\bibfnamefont{G.~F.} \bibnamefont{de~Arruda}}, \bibinfo{author}{\bibfnamefont{G.}~\bibnamefont{Petri}}, \bibnamefont{and} \bibinfo{author}{\bibfnamefont{Y.}~\bibnamefont{Moreno}}, \bibinfo{journal}{Phys. Rev. Research} \textbf{\bibinfo{volume}{2}}, \bibinfo{pages}{023032} (\bibinfo{year}{2020}).

\bibitem[{\citenamefont{Battiston et~al.}(2020)\citenamefont{Battiston, Cencetti, Iacopini, Latora, Lucas, Patania, Young, and Petri}}]{Battiston2021}
\bibinfo{author}{\bibfnamefont{F.}~\bibnamefont{Battiston}}, \bibinfo{author}{\bibfnamefont{G.}~\bibnamefont{Cencetti}}, \bibinfo{author}{\bibfnamefont{I.}~\bibnamefont{Iacopini}}, \bibinfo{author}{\bibfnamefont{V.}~\bibnamefont{Latora}}, \bibinfo{author}{\bibfnamefont{M.}~\bibnamefont{Lucas}}, \bibinfo{author}{\bibfnamefont{A.}~\bibnamefont{Patania}}, \bibinfo{author}{\bibfnamefont{J.-G.} \bibnamefont{Young}}, \bibnamefont{and} \bibinfo{author}{\bibfnamefont{G.}~\bibnamefont{Petri}}, \bibinfo{journal}{Physics Reports} \textbf{\bibinfo{volume}{874}}, \bibinfo{pages}{1} (\bibinfo{year}{2020}), ISSN \bibinfo{issn}{0370-1573}, \bibinfo{note}{networks beyond pairwise interactions: Structure and dynamics}.

\bibitem[{\citenamefont{Barrat et~al.}(2022)\citenamefont{Barrat, Ferraz~de Arruda, Iacopini, and Moreno}}]{barrat2021social}
\bibinfo{author}{\bibfnamefont{A.}~\bibnamefont{Barrat}}, \bibinfo{author}{\bibfnamefont{G.}~\bibnamefont{Ferraz~de Arruda}}, \bibinfo{author}{\bibfnamefont{I.}~\bibnamefont{Iacopini}}, \bibnamefont{and} \bibinfo{author}{\bibfnamefont{Y.}~\bibnamefont{Moreno}}, \emph{\bibinfo{title}{Social Contagion on Higher-Order Structures}} (\bibinfo{publisher}{Springer International Publishing}, \bibinfo{address}{Cham}, \bibinfo{year}{2022}), pp. \bibinfo{pages}{329--346}.

\bibitem[{\citenamefont{Battiston et~al.}(2021)\citenamefont{Battiston, Amico, Barrat, Bianconi, Ferraz~de Arruda, Franceschiello, Iacopini, K{\'e}fi, Latora, Moreno et~al.}}]{Battiston2021b}
\bibinfo{author}{\bibfnamefont{F.}~\bibnamefont{Battiston}}, \bibinfo{author}{\bibfnamefont{E.}~\bibnamefont{Amico}}, \bibinfo{author}{\bibfnamefont{A.}~\bibnamefont{Barrat}}, \bibinfo{author}{\bibfnamefont{G.}~\bibnamefont{Bianconi}}, \bibinfo{author}{\bibfnamefont{G.}~\bibnamefont{Ferraz~de Arruda}}, \bibinfo{author}{\bibfnamefont{B.}~\bibnamefont{Franceschiello}}, \bibinfo{author}{\bibfnamefont{I.}~\bibnamefont{Iacopini}}, \bibinfo{author}{\bibfnamefont{S.}~\bibnamefont{K{\'e}fi}}, \bibinfo{author}{\bibfnamefont{V.}~\bibnamefont{Latora}}, \bibinfo{author}{\bibfnamefont{Y.}~\bibnamefont{Moreno}}, \bibnamefont{et~al.}, \bibinfo{journal}{Nature Physics} \textbf{\bibinfo{volume}{17}}, \bibinfo{pages}{1093} (\bibinfo{year}{2021}), ISSN \bibinfo{issn}{1745-2481}.

\bibitem[{\citenamefont{Ferraz~de Arruda et~al.}(2023)\citenamefont{Ferraz~de Arruda, Petri, Rodriguez, and Moreno}}]{Arruda2023}
\bibinfo{author}{\bibfnamefont{G.}~\bibnamefont{Ferraz~de Arruda}}, \bibinfo{author}{\bibfnamefont{G.}~\bibnamefont{Petri}}, \bibinfo{author}{\bibfnamefont{P.~M.} \bibnamefont{Rodriguez}}, \bibnamefont{and} \bibinfo{author}{\bibfnamefont{Y.}~\bibnamefont{Moreno}}, \bibinfo{journal}{Nature Communications} \textbf{\bibinfo{volume}{14}}, \bibinfo{pages}{1375} (\bibinfo{year}{2023}), ISSN \bibinfo{issn}{2041-1723}.

\bibitem[{\citenamefont{Boccaletti et~al.}(2023)\citenamefont{Boccaletti, {De Lellis}, {del Genio}, Alfaro-Bittner, Criado, Jalan, and Romance}}]{Boccaletti2023}
\bibinfo{author}{\bibfnamefont{S.}~\bibnamefont{Boccaletti}}, \bibinfo{author}{\bibfnamefont{P.}~\bibnamefont{{De Lellis}}}, \bibinfo{author}{\bibfnamefont{C.}~\bibnamefont{{del Genio}}}, \bibinfo{author}{\bibfnamefont{K.}~\bibnamefont{Alfaro-Bittner}}, \bibinfo{author}{\bibfnamefont{R.}~\bibnamefont{Criado}}, \bibinfo{author}{\bibfnamefont{S.}~\bibnamefont{Jalan}}, \bibnamefont{and} \bibinfo{author}{\bibfnamefont{M.}~\bibnamefont{Romance}}, \bibinfo{journal}{Physics Reports} \textbf{\bibinfo{volume}{1018}}, \bibinfo{pages}{1} (\bibinfo{year}{2023}), ISSN \bibinfo{issn}{0370-1573}, \bibinfo{note}{the structure and dynamics of networks with higher order interactions}.

\bibitem[{\citenamefont{Alvarez-Rodriguez et~al.}(2021)\citenamefont{Alvarez-Rodriguez, Battiston, de~Arruda, Moreno, Perc, and Latora}}]{Alvarez-Rodriguez2021}
\bibinfo{author}{\bibfnamefont{U.}~\bibnamefont{Alvarez-Rodriguez}}, \bibinfo{author}{\bibfnamefont{F.}~\bibnamefont{Battiston}}, \bibinfo{author}{\bibfnamefont{G.~F.} \bibnamefont{de~Arruda}}, \bibinfo{author}{\bibfnamefont{Y.}~\bibnamefont{Moreno}}, \bibinfo{author}{\bibfnamefont{M.}~\bibnamefont{Perc}}, \bibnamefont{and} \bibinfo{author}{\bibfnamefont{V.}~\bibnamefont{Latora}}, \bibinfo{journal}{Nature Human Behaviour} \textbf{\bibinfo{volume}{5}}, \bibinfo{pages}{586} (\bibinfo{year}{2021}).

\bibitem[{\citenamefont{Bradde and Bianconi}(2009)}]{Bradde_2009}
\bibinfo{author}{\bibfnamefont{S.}~\bibnamefont{Bradde}} \bibnamefont{and} \bibinfo{author}{\bibfnamefont{G.}~\bibnamefont{Bianconi}}, \bibinfo{journal}{Journal of Statistical Mechanics: Theory and Experiment} \textbf{\bibinfo{volume}{2009}}, \bibinfo{pages}{P07028} (\bibinfo{year}{2009}).

\bibitem[{\citenamefont{Bianconi and Dorogovtsev}(2023)}]{bianconi2023theory}
\bibinfo{author}{\bibfnamefont{G.}~\bibnamefont{Bianconi}} \bibnamefont{and} \bibinfo{author}{\bibfnamefont{S.~N.} \bibnamefont{Dorogovtsev}}, \emph{\bibinfo{title}{The theory of percolation on hypergraphs}} (\bibinfo{year}{2023}), \eprint{2305.12297}.

\bibitem[{\citenamefont{Chitra and Raphael}(2019)}]{Chitra2019}
\bibinfo{author}{\bibfnamefont{U.}~\bibnamefont{Chitra}} \bibnamefont{and} \bibinfo{author}{\bibfnamefont{B.~J.} \bibnamefont{Raphael}}, in \emph{\bibinfo{booktitle}{Proceedings of the 36th International Conference on Machine Learning, {ICML} 2019, 9-15 June 2019, Long Beach, California, {USA}}}, edited by \bibinfo{editor}{\bibfnamefont{K.}~\bibnamefont{Chaudhuri}} \bibnamefont{and} \bibinfo{editor}{\bibfnamefont{R.}~\bibnamefont{Salakhutdinov}} (\bibinfo{publisher}{PMLR}, \bibinfo{address}{Long Beach, California, USA}, \bibinfo{year}{2019}), vol.~\bibinfo{volume}{97} of \emph{\bibinfo{series}{Proceedings of Machine Learning Research}}, pp. \bibinfo{pages}{1172--1181}.

\bibitem[{\citenamefont{Jamakovic and Van~Mieghem}(2008)}]{Jamakovic2008}
\bibinfo{author}{\bibfnamefont{A.}~\bibnamefont{Jamakovic}} \bibnamefont{and} \bibinfo{author}{\bibfnamefont{P.}~\bibnamefont{Van~Mieghem}}, in \emph{\bibinfo{booktitle}{NETWORKING 2008 Ad Hoc and Sensor Networks, Wireless Networks, Next Generation Internet}}, edited by \bibinfo{editor}{\bibfnamefont{A.}~\bibnamefont{Das}}, \bibinfo{editor}{\bibfnamefont{H.~K.} \bibnamefont{Pung}}, \bibinfo{editor}{\bibfnamefont{F.~B.~S.} \bibnamefont{Lee}}, \bibnamefont{and} \bibinfo{editor}{\bibfnamefont{L.~W.~C.} \bibnamefont{Wong}} (\bibinfo{publisher}{Springer Berlin Heidelberg}, \bibinfo{address}{Berlin, Heidelberg}, \bibinfo{year}{2008}), pp. \bibinfo{pages}{183--194}.

\bibitem[{\citenamefont{Cozzo et~al.}(2019)\citenamefont{Cozzo, de~Arruda, Rodrigues, and Moreno}}]{Cozzo2019}
\bibinfo{author}{\bibfnamefont{E.}~\bibnamefont{Cozzo}}, \bibinfo{author}{\bibfnamefont{G.~F.} \bibnamefont{de~Arruda}}, \bibinfo{author}{\bibfnamefont{F.~A.} \bibnamefont{Rodrigues}}, \bibnamefont{and} \bibinfo{author}{\bibfnamefont{Y.}~\bibnamefont{Moreno}}, \bibinfo{journal}{Phys. Rev. E} \textbf{\bibinfo{volume}{100}}, \bibinfo{pages}{012313} (\bibinfo{year}{2019}).

\bibitem[{APS()}]{APS_website}
\emph{\bibinfo{title}{{APS} data sets for research}}, \urlprefix\url{https://journals.aps.org/datasets}.

\bibitem[{\citenamefont{Sinha et~al.}(2015)\citenamefont{Sinha, Shen, Song, Ma, Eide, Hsu, and Wang}}]{sinha2015overview}
\bibinfo{author}{\bibfnamefont{A.}~\bibnamefont{Sinha}}, \bibinfo{author}{\bibfnamefont{Z.}~\bibnamefont{Shen}}, \bibinfo{author}{\bibfnamefont{Y.}~\bibnamefont{Song}}, \bibinfo{author}{\bibfnamefont{H.}~\bibnamefont{Ma}}, \bibinfo{author}{\bibfnamefont{D.}~\bibnamefont{Eide}}, \bibinfo{author}{\bibfnamefont{B.-J.} \bibnamefont{Hsu}}, \bibnamefont{and} \bibinfo{author}{\bibfnamefont{K.}~\bibnamefont{Wang}}, in \emph{\bibinfo{booktitle}{{Proceedings of the 24th International Conference on World Wide Web}}} (\bibinfo{year}{2015}), pp. \bibinfo{pages}{243--246}.

\bibitem[{\citenamefont{Feng et~al.}(2019)\citenamefont{Feng, You, Zhang, Ji, and Gao}}]{Feng2019}
\bibinfo{author}{\bibfnamefont{Y.}~\bibnamefont{Feng}}, \bibinfo{author}{\bibfnamefont{H.}~\bibnamefont{You}}, \bibinfo{author}{\bibfnamefont{Z.}~\bibnamefont{Zhang}}, \bibinfo{author}{\bibfnamefont{R.}~\bibnamefont{Ji}}, \bibnamefont{and} \bibinfo{author}{\bibfnamefont{Y.}~\bibnamefont{Gao}}, in \emph{\bibinfo{booktitle}{AAAI}} (\bibinfo{publisher}{AAAI Press}, \bibinfo{year}{2019}), pp. \bibinfo{pages}{3558--3565}.

\bibitem[{\citenamefont{Yadati et~al.}(2019)\citenamefont{Yadati, Nimishakavi, Yadav, Nitin, Louis, and Talukdar}}]{Yadati2019}
\bibinfo{author}{\bibfnamefont{N.}~\bibnamefont{Yadati}}, \bibinfo{author}{\bibfnamefont{M.}~\bibnamefont{Nimishakavi}}, \bibinfo{author}{\bibfnamefont{P.}~\bibnamefont{Yadav}}, \bibinfo{author}{\bibfnamefont{V.}~\bibnamefont{Nitin}}, \bibinfo{author}{\bibfnamefont{A.}~\bibnamefont{Louis}}, \bibnamefont{and} \bibinfo{author}{\bibfnamefont{P.}~\bibnamefont{Talukdar}}, \emph{\bibinfo{title}{HyperGCN: A New Method of Training Graph Convolutional Networks on Hypergraphs}} (\bibinfo{publisher}{Curran Associates Inc.}, \bibinfo{address}{Red Hook, NY, USA}, \bibinfo{year}{2019}).

\bibitem[{\citenamefont{Ji et~al.}(2020)\citenamefont{Ji, Feng, Ji, Zhao, Tang, and Gao}}]{Ji2020}
\bibinfo{author}{\bibfnamefont{S.}~\bibnamefont{Ji}}, \bibinfo{author}{\bibfnamefont{Y.}~\bibnamefont{Feng}}, \bibinfo{author}{\bibfnamefont{R.}~\bibnamefont{Ji}}, \bibinfo{author}{\bibfnamefont{X.}~\bibnamefont{Zhao}}, \bibinfo{author}{\bibfnamefont{W.}~\bibnamefont{Tang}}, \bibnamefont{and} \bibinfo{author}{\bibfnamefont{Y.}~\bibnamefont{Gao}}, in \emph{\bibinfo{booktitle}{Proceedings of the 26th ACM SIGKDD International Conference on Knowledge Discovery \& Data Mining}} (\bibinfo{publisher}{Association for Computing Machinery}, \bibinfo{address}{New York, NY, USA}, \bibinfo{year}{2020}), KDD '20, p. \bibinfo{pages}{2020–2029}, ISBN \bibinfo{isbn}{9781450379984}.

\bibitem[{\citenamefont{Dong et~al.}(2020)\citenamefont{Dong, Sawin, and Bengio}}]{HNHN2020}
\bibinfo{author}{\bibfnamefont{Y.}~\bibnamefont{Dong}}, \bibinfo{author}{\bibfnamefont{W.}~\bibnamefont{Sawin}}, \bibnamefont{and} \bibinfo{author}{\bibfnamefont{Y.}~\bibnamefont{Bengio}}, \bibinfo{journal}{ICML Graph Representation Learning and Beyond Workshop}  (\bibinfo{year}{2020}), \urlprefix\url{https://arxiv.org/abs/2006.12278}.

\bibitem[{\citenamefont{Huang and Yang}(2021)}]{ijcai21-UniGNN}
\bibinfo{author}{\bibfnamefont{J.}~\bibnamefont{Huang}} \bibnamefont{and} \bibinfo{author}{\bibfnamefont{J.}~\bibnamefont{Yang}}, in \emph{\bibinfo{booktitle}{Proceedings of the Thirtieth International Joint Conference on Artificial Intelligence, {IJCAI-21}}} (\bibinfo{year}{2021}).

\bibitem[{\citenamefont{Gao et~al.}(2023)\citenamefont{Gao, Feng, Ji, and Ji}}]{Gao2023}
\bibinfo{author}{\bibfnamefont{Y.}~\bibnamefont{Gao}}, \bibinfo{author}{\bibfnamefont{Y.}~\bibnamefont{Feng}}, \bibinfo{author}{\bibfnamefont{S.}~\bibnamefont{Ji}}, \bibnamefont{and} \bibinfo{author}{\bibfnamefont{R.}~\bibnamefont{Ji}}, \bibinfo{journal}{IEEE Transactions on Pattern Analysis and Machine Intelligence} \textbf{\bibinfo{volume}{45}}, \bibinfo{pages}{3181} (\bibinfo{year}{2023}).

\bibitem[{\citenamefont{Li et~al.}(2023)\citenamefont{Li, Zhang, Li, Zhang, and Yin}}]{Li2023}
\bibinfo{author}{\bibfnamefont{M.}~\bibnamefont{Li}}, \bibinfo{author}{\bibfnamefont{Y.}~\bibnamefont{Zhang}}, \bibinfo{author}{\bibfnamefont{X.}~\bibnamefont{Li}}, \bibinfo{author}{\bibfnamefont{Y.}~\bibnamefont{Zhang}}, \bibnamefont{and} \bibinfo{author}{\bibfnamefont{B.}~\bibnamefont{Yin}}, \bibinfo{journal}{ACM Trans. Knowl. Discov. Data} \textbf{\bibinfo{volume}{17}} (\bibinfo{year}{2023}), ISSN \bibinfo{issn}{1556-4681}.

\bibitem[{\citenamefont{Jost and Mulas}(2019)}]{Jost2019}
\bibinfo{author}{\bibfnamefont{J.}~\bibnamefont{Jost}} \bibnamefont{and} \bibinfo{author}{\bibfnamefont{R.}~\bibnamefont{Mulas}}, \bibinfo{journal}{Advances in Mathematics} \textbf{\bibinfo{volume}{351}}, \bibinfo{pages}{870} (\bibinfo{year}{2019}).

\bibitem[{\citenamefont{Mulas and Zhang}(2021)}]{Mulas2021}
\bibinfo{author}{\bibfnamefont{R.}~\bibnamefont{Mulas}} \bibnamefont{and} \bibinfo{author}{\bibfnamefont{D.}~\bibnamefont{Zhang}}, \bibinfo{journal}{Discrete Mathematics} \textbf{\bibinfo{volume}{344}}, \bibinfo{pages}{112372} (\bibinfo{year}{2021}), ISSN \bibinfo{issn}{0012-365X}.

\bibitem[{\citenamefont{Chung}(2005)}]{Chung2005}
\bibinfo{author}{\bibfnamefont{F.}~\bibnamefont{Chung}}, \bibinfo{journal}{Annals of Combinatorics} \textbf{\bibinfo{volume}{9}}, \bibinfo{pages}{1} (\bibinfo{year}{2005}).

\bibitem[{\citenamefont{Rehman et~al.}(2021)\citenamefont{Rehman, Zamli, Almutairi, Chiroma, Aamir, Kader, and Nawi}}]{Rehman2021}
\bibinfo{author}{\bibfnamefont{M.~Z.} \bibnamefont{Rehman}}, \bibinfo{author}{\bibfnamefont{K.~Z.} \bibnamefont{Zamli}}, \bibinfo{author}{\bibfnamefont{M.}~\bibnamefont{Almutairi}}, \bibinfo{author}{\bibfnamefont{H.}~\bibnamefont{Chiroma}}, \bibinfo{author}{\bibfnamefont{M.}~\bibnamefont{Aamir}}, \bibinfo{author}{\bibfnamefont{M.~A.} \bibnamefont{Kader}}, \bibnamefont{and} \bibinfo{author}{\bibfnamefont{N.~M.} \bibnamefont{Nawi}}, \bibinfo{journal}{PLOS ONE} \textbf{\bibinfo{volume}{16}}, \bibinfo{pages}{1} (\bibinfo{year}{2021}).

\bibitem[{\citenamefont{Huang et~al.}(2013)\citenamefont{Huang, Vodenska, Havlin, and Stanley}}]{Huang2013}
\bibinfo{author}{\bibfnamefont{X.}~\bibnamefont{Huang}}, \bibinfo{author}{\bibfnamefont{I.}~\bibnamefont{Vodenska}}, \bibinfo{author}{\bibfnamefont{S.}~\bibnamefont{Havlin}}, \bibnamefont{and} \bibinfo{author}{\bibfnamefont{H.~E.} \bibnamefont{Stanley}}, \bibinfo{journal}{Scientific Reports} \textbf{\bibinfo{volume}{3}}, \bibinfo{pages}{1219} (\bibinfo{year}{2013}), ISSN \bibinfo{issn}{2045-2322}.

\bibitem[{\citenamefont{Caccioli et~al.}(2014)\citenamefont{Caccioli, Shrestha, Moore, and Farmer}}]{Caccioli2014}
\bibinfo{author}{\bibfnamefont{F.}~\bibnamefont{Caccioli}}, \bibinfo{author}{\bibfnamefont{M.}~\bibnamefont{Shrestha}}, \bibinfo{author}{\bibfnamefont{C.}~\bibnamefont{Moore}}, \bibnamefont{and} \bibinfo{author}{\bibfnamefont{J.~D.} \bibnamefont{Farmer}}, \bibinfo{journal}{Journal of Banking \& Finance} \textbf{\bibinfo{volume}{46}}, \bibinfo{pages}{233} (\bibinfo{year}{2014}), ISSN \bibinfo{issn}{0378-4266}.

\bibitem[{\citenamefont{Liemhetcharat and Veloso}(2012)}]{Liemhetcharat2012}
\bibinfo{author}{\bibfnamefont{S.}~\bibnamefont{Liemhetcharat}} \bibnamefont{and} \bibinfo{author}{\bibfnamefont{M.}~\bibnamefont{Veloso}}, in \emph{\bibinfo{booktitle}{Proceedings of the 11th International Conference on Autonomous Agents and Multiagent Systems - Volume 1}} (\bibinfo{publisher}{International Foundation for Autonomous Agents and Multiagent Systems}, \bibinfo{address}{Richland, SC}, \bibinfo{year}{2012}), AAMAS '12, p. \bibinfo{pages}{365–374}, ISBN \bibinfo{isbn}{0981738117}.

\bibitem[{\citenamefont{Liemhetcharat and Veloso}(2014)}]{Liemhetcharat2014}
\bibinfo{author}{\bibfnamefont{S.}~\bibnamefont{Liemhetcharat}} \bibnamefont{and} \bibinfo{author}{\bibfnamefont{M.}~\bibnamefont{Veloso}}, \bibinfo{journal}{Artificial Intelligence} \textbf{\bibinfo{volume}{208}}, \bibinfo{pages}{41} (\bibinfo{year}{2014}), ISSN \bibinfo{issn}{0004-3702}.

\bibitem[{\citenamefont{Mabry et~al.}(2020)\citenamefont{Mabry, Yan, Pentchev, {Van Rennes}, McGavin, and Wittenberg}}]{mabry2020cadre}
\bibinfo{author}{\bibfnamefont{P.~L.} \bibnamefont{Mabry}}, \bibinfo{author}{\bibfnamefont{X.}~\bibnamefont{Yan}}, \bibinfo{author}{\bibfnamefont{V.}~\bibnamefont{Pentchev}}, \bibinfo{author}{\bibfnamefont{R.}~\bibnamefont{{Van Rennes}}}, \bibinfo{author}{\bibfnamefont{S.~H.} \bibnamefont{McGavin}}, \bibnamefont{and} \bibinfo{author}{\bibfnamefont{J.~V.} \bibnamefont{Wittenberg}}, \bibinfo{journal}{Frontiers in Big Data} \textbf{\bibinfo{volume}{3}}, \bibinfo{pages}{42} (\bibinfo{year}{2020}).

\bibitem[{\citenamefont{Kingma and Ba}(2014)}]{kingma2014adam}
\bibinfo{author}{\bibfnamefont{D.~P.} \bibnamefont{Kingma}} \bibnamefont{and} \bibinfo{author}{\bibfnamefont{J.}~\bibnamefont{Ba}}, \bibinfo{journal}{arXiv preprint arXiv:1412.6980}  (\bibinfo{year}{2014}).

\bibitem[{\citenamefont{Schuetz et~al.}(2022)\citenamefont{Schuetz, Brubaker, and Katzgraber}}]{schuetz2022combinatorial}
\bibinfo{author}{\bibfnamefont{M.~J.} \bibnamefont{Schuetz}}, \bibinfo{author}{\bibfnamefont{J.~K.} \bibnamefont{Brubaker}}, \bibnamefont{and} \bibinfo{author}{\bibfnamefont{H.~G.} \bibnamefont{Katzgraber}}, \bibinfo{journal}{Nature Machine Intelligence} \textbf{\bibinfo{volume}{4}}, \bibinfo{pages}{367} (\bibinfo{year}{2022}).

\bibitem[{\citenamefont{Heydaribeni et~al.}(2023)\citenamefont{Heydaribeni, Zhan, Zhang, Eliassi-Rad, and Koushanfar}}]{heydaribeni2023hypop}
\bibinfo{author}{\bibfnamefont{N.}~\bibnamefont{Heydaribeni}}, \bibinfo{author}{\bibfnamefont{X.}~\bibnamefont{Zhan}}, \bibinfo{author}{\bibfnamefont{R.}~\bibnamefont{Zhang}}, \bibinfo{author}{\bibfnamefont{T.}~\bibnamefont{Eliassi-Rad}}, \bibnamefont{and} \bibinfo{author}{\bibfnamefont{F.}~\bibnamefont{Koushanfar}}, \bibinfo{journal}{arXiv preprint arXiv:2311.09375}  (\bibinfo{year}{2023}).

\bibitem[{\citenamefont{Cappart et~al.}(2023)\citenamefont{Cappart, Ch{\'e}telat, Khalil, Lodi, Morris, and Velickovic}}]{cappart2023combinatorial}
\bibinfo{author}{\bibfnamefont{Q.}~\bibnamefont{Cappart}}, \bibinfo{author}{\bibfnamefont{D.}~\bibnamefont{Ch{\'e}telat}}, \bibinfo{author}{\bibfnamefont{E.~B.} \bibnamefont{Khalil}}, \bibinfo{author}{\bibfnamefont{A.}~\bibnamefont{Lodi}}, \bibinfo{author}{\bibfnamefont{C.}~\bibnamefont{Morris}}, \bibnamefont{and} \bibinfo{author}{\bibfnamefont{P.}~\bibnamefont{Velickovic}}, \bibinfo{journal}{J. Mach. Learn. Res.} \textbf{\bibinfo{volume}{24}}, \bibinfo{pages}{130} (\bibinfo{year}{2023}).

\bibitem[{Gur()}]{Gurobi}
\emph{\bibinfo{title}{Gurobi optimization}}, \urlprefix\url{https://www.gurobi.com}.

\bibitem[{\citenamefont{Javaheripi et~al.}(2020)\citenamefont{Javaheripi, Samragh, Javidi, and Koushanfar}}]{javaheripi2020adans}
\bibinfo{author}{\bibfnamefont{M.}~\bibnamefont{Javaheripi}}, \bibinfo{author}{\bibfnamefont{M.}~\bibnamefont{Samragh}}, \bibinfo{author}{\bibfnamefont{T.}~\bibnamefont{Javidi}}, \bibnamefont{and} \bibinfo{author}{\bibfnamefont{F.}~\bibnamefont{Koushanfar}}, \bibinfo{journal}{IEEE Journal of Selected Topics in Signal Processing} \textbf{\bibinfo{volume}{14}}, \bibinfo{pages}{750} (\bibinfo{year}{2020}).

\end{thebibliography}

\end{document}